\documentclass[12pt]{article}
\usepackage[utf8]{inputenc}
\usepackage{amsmath, amssymb, amsthm, bbm, mathrsfs}
\usepackage{geometry}
\usepackage{natbib}
\usepackage{verbatim}
\usepackage{multirow}
\usepackage{booktabs, array, threeparttable}
\usepackage{threeparttable}
\usepackage{rotating}
\usepackage{graphicx}
\usepackage{subfigure}
\usepackage{float}
\usepackage{caption}
\usepackage{amsfonts}
\usepackage{xr}
\usepackage{color}
\usepackage{enumerate}
\usepackage{epstopdf}
\usepackage{threeparttable}
\usepackage{siunitx}
\usepackage{multirow}
\usepackage[colorlinks=true, allcolors=blue]{hyperref}
\usepackage{makecell}
\allowdisplaybreaks
\usepackage[linesnumbered,ruled,vlined]{algorithm2e}
\SetKwInput{KwInput}{Input}                
\SetKwInput{KwOutput}{Output}  
\SetKwInput{KwSet}{Set}            
\SetKwProg{Fn}{Function}{:}{}
\usepackage{geometry}
\graphicspath{{./figs/}}

\newcommand{\T}{{\top}}

\newcommand{\cT}{{\mathcal{T}}}

\newcommand{\cI}{{\mathcal{I}}}

\newcommand{\bbE}{{\mathbb{E}}}

\newcommand{\commenting}[1]{}

\newtheorem{theorem}{Theorem}
\newtheorem{lemma}{Lemma}
\newtheorem{remark}{Remark}
\newtheorem{corollary}{Corollary}
\newtheorem{assumption}{Assumption}
\newtheorem{example}{Example}

\newtheorem{proposition}{Proposition}
\newtheorem{definition}{Definition}

\def\art{ART}
\def\s{\mathbb{S}}
\def\S{S}
\def\a{\mathbb{A}}
\def\g{\mathbb{G}}
\def\A{T}
\def\dev{\mathbb{D}}
\def\lab{\mathbb{C}}
\def\thresh{t}
\def\sl{s_\ell}
\def\el{e_\ell}

\title{\art: Distribution-Free and Model-Agnostic Changepoint Detection with Finite-Sample Guarantees}
\date{}
\author{Xiaolong Cui, Haoyu Geng, Guanghui Wang, Zhaojun Wang and Changliang Zou\\
{\small\it School of Statistics and Data Science, Nankai University, China}}

\begin{document}

\maketitle

\begin{abstract}
\baselineskip 20pt
We introduce ART, a distribution-free and model-agnostic framework for changepoint detection that provides finite-sample guarantees. ART transforms independent observations into real-valued scores via a symmetric function, ensuring exchangeability in the absence of changepoints. These scores are then ranked and aggregated to detect distributional changes. The resulting test offers exact Type-I error control, agnostic to specific distributional or model assumptions. Moreover, ART seamlessly extends to multi-scale settings, enabling robust multiple changepoint estimation and post-detection inference with finite-sample error rate control. By locally ranking the scores and performing aggregations across multiple prespecified intervals, ART identifies changepoint intervals and refines subsequent inference while maintaining its distribution-free and model-agnostic nature. This adaptability makes ART as a reliable and versatile tool for modern changepoint analysis, particularly in high-dimensional data contexts and applications leveraging machine learning methods.

\end{abstract}

\section{Introduction}\label{sec:introduction}

Changepoint analysis focuses on identifying abrupt changes in data sequences, a common occurrence when an underlying process evolves over time or across other indexing domains. Such changes arise in diverse settings, including mean-level shifts in financial series, disruptions in genomic profiles, and anomalies in streaming sensor data.

Over the past few decades, extensive research has examined whether, and where, one or more changepoints occur. Early efforts explored specific finite-dimensional parametric models \citep{csorgo1997limit,killick2012optimal,truong2020selective} and nonparametric models \citep{matteson2014nonparametric,zou2014nonparametric,lung2015homogeneity,chen2015graph,arlot2019kernel}. More recent studies address high-dimensional scenarios, investigating changes in means \citep{MR2652280,MR3405600,cho2015multiple,wang2018high,MR3845009,liu2020unified,yu2021finite,zhang2022adaptive,wang2023computationally}, regression coefficients \citep{lee2016lasso,leonardi2016computationally,kaul2019efficient,wang2021statistically,xu2024change}, and general distributional shifts \citep{londschien2023random,10.1093/jrsssb/qkae004}. Concurrently, a growing literature emphasizes uncertainty quantification of detected changepoints \citep{MR3222810,frick2014multiscale,fang2020segmentation,chen_data-driven_2021,fryzlewicz2024narrowest,fryzlewicz2024robust} and post-detection inference \citep{hyun2018exact,hyun2021post,duy2020computing,jewell2022testing,carrington2025improving}. These developments are crucial, given that perfect model recovery is often elusive in practice.

Most changepoint testing procedures rely on asymptotic null distributions to obtain critical values. These tests often guide changepoint estimation methods such as binary segmentation \citep{fryzlewicz2014wild,baranowski2019narrowest,kovacs2023seeded,yu2021finite}, moving windows \citep{niu2012screening,eichinger2018mosum}, and scanning-based algorithms \citep{MR3076173}. In addition, the critical values or corresponding p-values help control global error rates for uncertainty quantification \citep{frick2014multiscale,fang2020segmentation,fryzlewicz2024narrowest}. However, convergence to asymptotic null distributions can often be slow \citep{csorgo1997limit}, and their accuracies usually depend on stringent assumptions about the data distribution and model form. These approximations become even more fragile in high-dimensional or complex changepoint settings, especially when flexible estimation procedures are employed \citep{zhao2024optimal,londschien2023random}. Collectively, these limitations can erode the reliability of changepoint testing, estimation, and post-detection inference in finite-sample scenarios.

To reduce reliance on specific parametric models and distributional assumptions, various nonparametric approaches have been developed. In the univariate setting, rank-based distribution-free tests \citep{pettitt1979non, lombard1987rank, MR1997030} and permutation tests \citep{antoch2001permutation} are well established. In the multivariate context, \cite{chen2015graph} presented a distribution-free graph-based test, while \cite{lung2015homogeneity} enhanced rank-based methods by aggregating component-wise rank statistics using the $L_2$-norm, both offering asymptotic guarantees. For high-dimensional data, \cite{yu2022robust} employed the $L_\infty$-norm with critical values derived from bootstraps. More recently, permutation-based techniques have been heuristically incorporated into distance-based \citep{matteson2014nonparametric} and classifier-based \citep{londschien2023random} methods for multivariate data. Despite these advances, adapting nonparametric changepoint detection methods to supervised scenarios---such as detecting changes in regression coefficients---remains challenging, especially in flexible, high- or infinite-dimensional settings.

The overview highlights a crucial, yet largely unresolved question in changepoint analysis: Is it feasible to devise an exact, distribution-free test that accommodates a broad class of changepoint models under minimal assumptions on the underlying data? Furthermore, can such a framework deliver effective changepoint localization and facilitate reliable subsequent inference?

\subsection{Our contributions}

In this work, we introduce a simple yet effective framework, \underline{A}ggregation based on \underline{R}anks of \underline{T}ransformed sequences (abbreviated as \art), for changepoint detection and inference. Our methodology begins by transforming the original observations $\{Z_i:i\in[n]\}$---each potentially residing in an arbitrary space---into real-valued {scores} $\{\S_i:i\in[n]\}$, where $[n]=\{1,\ldots,n\}$. This transformation is \textit{symmetric} in the sense that it does not depend on the order of the observations, thereby rendering the scores \textit{exchangeable} in the absence of changepoints. A central component of the \art\ test is the ranking of these scores, where under the null hypothesis, the ranks $\{R_i:i\in[n]\}$ are uniformly distributed over all permutations of $[n]$, independent of the data's underlying distribution. We then apply an aggregation function $\a$ to these ranks, yielding a test statistic $\a(R_1,\ldots,R_n)$, with larger values indicating a rejection of the null hypothesis. Thanks to the distribution-free nature of the ranks, the null distribution of this statistic can be determined exactly. \art\ possesses several key features for changepoint testing:
\begin{itemize}
\item \textbf{Assumption-lean and model-agnostic}: \art\ imposes minimal conditions on the data-generating process (namely, independence) and applies broadly to diverse changepoint models and parameters of interest.
\item \textbf{Distribution-free and exact}: The null distribution relies solely on the uniform distribution over all permutations of $[n]$. Consequently, the \art\ test is exact, ensuring valid control of the Type-I error under finite-sample scenarios.
\end{itemize}
\noindent We further outline general recipes for creating transformed scores---e.g., through \textit{deviance} and \textit{clustering} transformations---and describe appropriate aggregation functions, including \textit{rank CUSUM} and \textit{nonparametric likelihood} aggregations. These ideas accommodate high- or infinite-dimensional parameters estimated via flexible statistical/machine learning methods, where designing valid tests can be difficult.

\begin{figure}[!ht]
\centering
\includegraphics[scale=0.3]{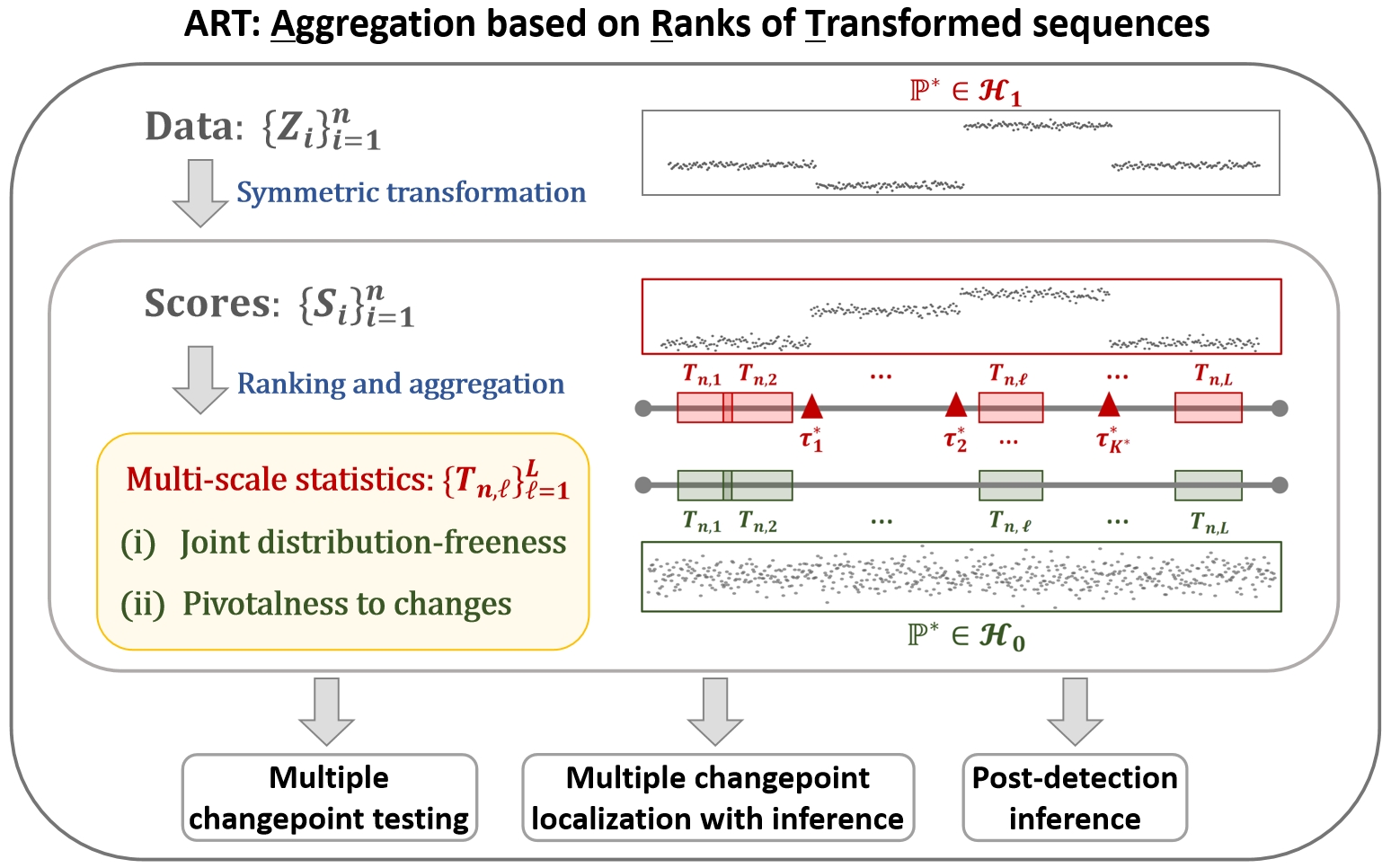}
\caption{A flowchart illustrating the procedure, key properties, and application scenarios of \art. Here, $\mathbb{P}^*\in\mathcal{H}_1$ characterizes the actual data distribution with changepoints $\{\tau^*_k\}_{k\in[K^*]}$, while $\mathbb{P}^*\in\mathcal{H}_0$ represents a hypothetical context without any changepoints; see Section \ref{subsec:model}.}
\label{fig:flowchart}
\end{figure}

More importantly, \art\ ingeniously extends to multi-scale settings, enabling multiple changepoint estimation and post-detection inference. Specifically, by locally ranking the scores and performing aggregations over multiple prespecified intervals $\{\mathcal{I}_\ell:\ell\in[L]\}$, we form an array of statistics $\A_{n,\ell}=\a(\{R_{i,\mathcal{I}_\ell}\}_{i\in\mathcal{I}_\ell})$ for $\ell\in[L]$, where $R_{i,\mathcal{I}}$ denotes the rank of $\S_i$ among $\{\S_j:j\in\mathcal{I}\}$. Despite their shared dependence on the same underlying scores, these statistics exhibit two critical properties: (i) They are jointly distribution-free if the entire dataset contains no changepoints, and (ii) Even if some intervals contain true changepoints, the joint distribution of any sub-collection of intervals that lie within homogeneous segments is the same as it would be in the absence of any changepoints, even though these intervals may individually exhibit distinct distributions. We refer to property (ii) as \textit{pivotalness to changes}. These properties are crucial in demonstrating that \art\ retains its distribution-free and model-agnostic attributes, enabling finite-sample control over certain global error rates. A schematic of the procedure and key use-cases is presented in Figure~\ref{fig:flowchart}.

\subsection{Related literature}

The \art\ test is distribution-free and exact, properties shared by classical nonparametric methods such as permutation tests \citep{MR4489085} and rank tests \citep{lehmann1975nonparametrics}. A permutation test involves a data-dependent critical value for a test statistic, where the null distribution remains invariant under permutations of the data. In contrast, rank tests typically yield a data-independent, exactly determinable critical value, as the null distribution is fully specified and does not depend on the data. Both approaches have a long history in changepoint testing, although much of the existing literature has focused on large-sample considerations \citep{pettitt1979non, MR1997030, antoch2001permutation, chen2015graph}. In this work, we integrate rank-based methods with recent advancements in randomized permutation techniques \citep{vovk2003testing, hemerik2018exact}, ensuring both exactness and computational feasibility for changepoint detection. Moreover, our transformation framework extends beyond univariate data, enabling the detection of general distributional changes or specific parametric shifts.

The model-agnostic and distribution-free features of \art, along with the use of transformed scores, resonate with those employed in conformal prediction \citep{vovk2005algorithmic, lei2018distribution, angelopoulos2024theoretical}. Both approaches are grounded in the principle of exchangeability. While conformal prediction focuses on quantifying prediction uncertainty for independent and identically distributed (iid) data, \art\ aims to control error rates associated with changepoint detection which distinguishes our transformation objectives from existing works (see Section \ref{subsec:trans}). Surprisingly, despite the non-iid nature of data in changepoint analysis, the combination of exchangeability-based rank statistics with multi-scale aggregations provide finite-sample guarantees.

Expanding on the ``pivotalness to changes'' property, \art\ integrates seamlessly with contemporary developments in uncertainty quantification for changepoint analysis \citep{fryzlewicz2024narrowest,jia2024tune}. This integration not only strengthens the contributions of these methods but also capitalizes on \art's inherent attributes of being distribution-free, model-agnostic, and capable of handling finite samples effectively.

\subsection{Structure and notations}

The remainder of this paper is structured as follows. Section \ref{sec:art} outlines the main ideas behind the \art\ framework, focusing on the notions of symmetric transformation, ranking and aggregations, and their multi-scale extensions. Section \ref{sec:MCP} delves into its application in multiple changepoint testing, localization with inference, and post-detection inference, particularly emphasizing the error rate control throughout these procedures. Simulation studies and real-data analyses are detailed in Section \ref{sec:numerical}. Section \ref{sec:conclusion} concludes the paper. All theoretical proofs, together with additional algorithmic and numerical details, are provided in the supplementary material.

\textbf{Notations}: The indicator function is denoted by $\mathbbm{1}\{\cdot\}$. The cardinality of set $A$ is given by $|A|$. Let $\|\cdot\|_2$ and $\|\cdot\|_{\infty}$ denote the $L_2$- and $L_{\infty}$-norms, respectively. For any $x\in \mathbb{R}$, let $\lfloor x \rfloor$ denote the greatest integer no larger than $x$. For random variables $X$ and $Y$, $X\overset{d}{=}Y$ indicates equality in distribution. The uniform distribution over $[0,1]$ is denoted by $\mathcal{U}(0,1)$, and $\mathcal{U}(\Pi_n)$ represents the uniform distribution over all permutations of $[n]$, each permutation $\pi$ having an equal probability of $1/n!$.

\section{\art\ methodology and key principles}\label{sec:art}

\subsection{Data and changepoint model}\label{subsec:model}

Consider a sequence of independent observations $\mathcal{D}=(Z_1,\ldots,Z_n)$, where each observation takes values in a space $\mathcal{Z}$. Let $\mathbb{P}^*$ denote the true joint distribution of the data.

Suppose that there are $K^*\ge 0$ changepoints at locations $0\equiv\tau^*_0<\tau^*_1<\cdots<\tau^*_{K^*}<\tau^*_{K^*+1}\equiv n$, partitioning the sequence into $K^*+1$ segments. Within each segment $k\in[K^*+1]$, the observations $\{Z_i:i\in(\tau^*_{k-1},\tau^*_k]\}$ are identically distributed according to a distribution $P^*_k$, with $P^*_{k+1}\ne P^*_k$ for $k\in[K^*]$. In situations where distributional changes can be described by a specific parameter, we express $P^*_k=P_{\theta^*_k}$, where $\theta^*_{k+1}\ne\theta^*_k$ for $k\in[K^*]$. Specific examples include:
\begin{example}[Changes in means]\label{ex:mean}
Let $\mathcal{Z}=\mathbb{R}^d$, and $Z_i=\theta^*_k+\varepsilon_i$ for $i\in(\tau^*_{k-1},\tau^*_k]$, where $\varepsilon_i$ are iid from $P_\epsilon$. Here, $P_{\theta^*_k}(z)=P_\epsilon(z-\theta^*_k)$.
\end{example}
\begin{example}[Changes in regression coefficients]\label{ex:reg}
Let $z=(y,x)\in\mathcal{Z}=\mathbb{R}\times\mathbb{R}^d$, and $y_i=x_i^\top\theta^*_k+\varepsilon_i$ for $i\in(\tau^*_{k-1},\tau^*_k]$, where the covariates $x_i$ are iid from $P_x$, the noises $\varepsilon_i$ are iid from $P_\epsilon$, and the covariates and noises are independent. Here, $P_{\theta^*_k}(z)=P_x\cdot P_\epsilon(y-x^\top\theta^*_k)$.
\end{example}

Let $\mathcal{P}$ denote the set of all distributions over $\mathcal{Z}$. Define $\mathcal{H}_0=\{P_1^n:P_1\in\mathcal{P}\}$ as the set of distributions in the absence of any changepoints (i.e., $K^*=0$). Conversely, define $\mathcal{H}_1=\{\prod_{k\in[K^*+1]}P_k^{\tau^*_k-\tau^*_{k-1}}:P_k\in\mathcal{P},P_{k+1}\ne P_k\text{ for }k\in[K^*]\}$ for distributions with $K^*>0$ changepoints at specified locations.

\subsection{Transformation}\label{subsec:trans}

Our methodology begins by transforming the observations into \textit{scores} $\S_i=\s(Z_i;\mathcal{D})$, where $\s$ is a \textit{symmetric} transformation function, defined as follows:
\begin{definition}[Symmetric transformation]\label{def:symmetric_trans}
A transformation function $\s:\mathcal{Z}\times\mathcal{Z}^n\rightarrow\mathbb{R}$ is symmetric if for any $z\in\mathcal{Z}$ and any permutation $\pi$ on $[n]$, $\s(z;\mathcal{D})=\s(z;\mathcal{D}_\pi)$, where $\mathcal{D}_\pi=(Z_{\pi(1)},\ldots,Z_{\pi(n)})$.
\end{definition}

This transformation is designed to generate scores that can reflect potential changepoints within the data. We introduce two general transformation approaches:
\begin{itemize}
\item \textit{Deviance transformation}: $\s(z;\mathcal{D})=\dev(z;\widehat{f}_\mathcal{D})$, which quantifies the deviation of an observation $z$ from a baseline model $\widehat{f}_\mathcal{D}$ trained on $\mathcal{D}$. We require $\widehat{f}_\mathcal{D}=\widehat{f}_{\mathcal{D}_\pi}$ for any permutation $\pi$ on $[n]$, ensuring the symmetry of $\s(z;\mathcal{D})$. For instance, in Example \ref{ex:reg}, a practical choice is $\dev(z;\widehat{f}_\mathcal{D})=(y-x^\top\widehat{\theta}_\mathcal{D})^2$, where $\widehat{\theta}_\mathcal{D}$ is a prespecified or estimated regression coefficients. In Example \ref{ex:mean}, one can use $\dev(z;\widehat{f}_\mathcal{D})=-{p}_{\widehat{\theta}_\mathcal{D}}(z)$, where ${p}_{\theta}(z)$ represents a predefined parametric density ${p}_{\theta}(\cdot)$ evaluated at $z$ and  $\widehat{\theta}_\mathcal{D}$ is a prespecified or estimated parameter.

This approach is conceptually similar to the conformal score function used in conformal prediction \citep[e.g.,][]{angelopoulos2024theoretical}, where each score quantifies the discrepancy between the true response (or a hypothetical one) and the model's prediction. We anticipate that changes in the original data will manifest as corresponding changes in the transformed scores, whether in location or distribution.

\item \textit{Clustering transformation}: $\s(z;\mathcal{D})=\lab(z;\widehat{f}_\mathcal{D})$, which employs a clustering algorithm $\widehat{f}_\mathcal{D}$ to assign a unique integer label to each observation $z$. Again, we require $\widehat{f}_\mathcal{D}=\widehat{f}_{\mathcal{D}_\pi}$ for any permutation $\pi$ on $[n]$. Suitable clustering algorithms include $K$-means, hierarchical clustering, density-based clustering, and spectral clustering. When specifying the number of clusters, strategies that are invariant to the order of data, such as the heuristic elbow method, the gap statistic \citep{tibshirani2001estimating}, or certain information criteria, should be employed.
\end{itemize}

Further details and examples are provided in Table \ref{tab:smry} in Section \ref{sec:numerical}. For changepoint analysis involving high-dimensional and non-Euclidean data \citep{MR4567802,londschien2023random,10.1093/jrsssb/qkae004}, variable selection and dimension reduction techniques that remain invariant to data order are applicable, prior to any transformation. A detailed transformation approach employing deep embedded clustering \citep{xie2016unsupervised} for such data is provided in Algorithm \ref{alg:deepembedding} in the supplementary material.

\subsection{Ranking and aggregation}\label{subsec:rank_and_agg}

We outline our methodology for detecting the presence of changepoints. Formally, we test the null hypothesis $H_0:\mathbb{P}^*\in\mathcal{H}_0$ against the alternative $H_1:\mathbb{P}^*\in\mathcal{H}_1$. Let $\Pr$ be the probability evaluated considering all random quantities, and $\Pr_{H_0}$ evaluates $\Pr$ under $H_0$.

Under $H_0$, the symmetric transformation renders the scores \textit{exchangeable}:
\[
(\S_1,\ldots,\S_n)\overset{d}{=}(\S_{\pi(1)},\ldots,\S_{\pi(n)}),
\]
for any permutation $\pi$ on $[n]$. A key component of our test involves the \textit{ranking} of these scores. The rank of each score $\S_i$ for $i\in[n]$ among $\{\S_j:j\in[n]\}$ is defined as:
\[
R_i\equiv R_{i,[n]}=|\{j\in[n]:\S_j\le\S_i\}|.
\]
To break ties in the ranks, particularly common with clustering transformations, we modify the scores to $\S_i+\epsilon e_i$, where $\epsilon>0$ is a small constant (e.g., $\epsilon=10^{-6}$) and $\{e_i:i\in[n]\}$ are iid $\mathcal{N}(0,1)$ random variables. Under $H_0$, $(R_1,\ldots,R_n)$ follows the uniform distribution over all permutations of $[n]$, namely $\mathcal{U}(\Pi_n)$. Our test \textit{aggregates} these ranks using a function $\a:[n]^n\rightarrow\mathbb{R}$, with larger values of the test statistic $\A_n=\a(R_1,\ldots,R_n)$ indicating evidence against $H_0$. This method is referred to as ``\underline{A}ggregation based on \underline{R}anks of \underline{T}ransformed sequences'' (\art).

We present two practical aggregation functions:
\begin{itemize}
\item \textit{Rank cumulative sum (CUSUM) aggregation}:
\begin{equation*}
\a(R_1,\ldots,R_n)=n^{-3/2}\sup_{1\le t<n}\left\vert\sum_{i=1}^t(R_i-\bar{R}_n)\right\vert,    
\end{equation*}
where $\bar{R}_n=n^{-1}\sum_{i=1}^nR_i=(n+1)/2$. This aggregation is equivalent to computing the maximum of the sequence of the absolute vales of Wilcoxon-Mann-Whitney two-sample statistics $n^{-3/2}\sum_{i=1}^t\sum_{j=t+1}^n\{\mathbbm{1}(\S_j\le\S_i)-1/2\}$ over $1\le t<n$. For observations $Z_i\in\mathbb{R}$ and scores $S_i=Z_i$, this test statistic is identical to the one proposed by  \cite{pettitt1979non}.
\item \textit{Nonparametric likelihood aggregation}:
\begin{equation*}
\a(R_1,\ldots,R_n)=\sup_{1\le t<n}2\int_{s\in\mathbb{R}}\Lambda([t],[n]\backslash[t];s)\widehat{F}_{[n]}(s)^{-1}\{1-\widehat{F}_{[n]}(s)\}^{-1}d\widehat{F}_{[n]}(s),
\end{equation*}
where $\widehat{F}_\mathcal{I}(s)=(|\mathcal{I}|+1)^{-1}\{\sum_{i\in\mathcal{I}}\mathbbm{1}(\S_i\le s)+0.5\}$ denotes the empirical distribution function of the scores $\{\S_i:i\in\mathcal{I}\}$, adjusted for continuity, and
\begin{equation*}
\begin{aligned}
\Lambda(\mathcal{I}_1,\mathcal{I}_2;s) = |\mathcal{I}_1|&\Big[\widehat{F}_{\mathcal{I}_1}(s)\log\frac{\widehat{F}_{\mathcal{I}_1}(s)}{\widehat{F}_{\mathcal{I}}(s)} + \{1-\widehat{F}_{\mathcal{I}_1}(s)\}\log\frac{1-\widehat{F}_{\mathcal{I}_1}(s)}{1-\widehat{F}_{\mathcal{I}}(s)}\Big]\\
&+|\mathcal{I}_2|\Big[\widehat{F}_{\mathcal{I}_2}(s)\log\frac{\widehat{F}_{\mathcal{I}_2}(s)}{\widehat{F}_{\mathcal{I}}(s)} + \{1-\widehat{F}_{\mathcal{I}_2}(s)\}\log\frac{1-\widehat{F}_{\mathcal{I}_2}(s)}{1-\widehat{F}_{\mathcal{I}}(s)}\Big]
\end{aligned}
\end{equation*}
represents the nonparametric likelihood ratio statistic for evaluating distributional equality of two samples $\{\S_i:i\in\mathcal{I}_1\}$ and $\{\S_i:i\in\mathcal{I}_2\}$ \citep{zhang2006powerful}, with $\mathcal{I}=\mathcal{I}_1\cup\mathcal{I}_2$. 
\end{itemize}

For independent observations $Z_i\in\mathbb{R}$ and scores $S_i=Z_i$, the rank CUSUM aggregation is renowned for its effectiveness in detecting locational changes within the scores. To capture other types of changes such as scale shifts, extensions employing linear rank statistics can be developed \citep{lombard1987rank}. Alternatively, we suggest the nonparametric likelihood aggregation, which is advantageous for detecting omnibus distributional changes. These aggregations rely solely on the ranks of the scores, ensuring their exact distribution-freeness under $H_0$. This property holds true for exchangeable scores in our scenarios, which are transformed from general objects $Z_i\in\mathcal{Z}$ and may be highly dependent.

The corresponding null distribution or p-value can be determined exactly through the enumeration of all $n!$ permutations, which, however, becomes computationally prohibitive for large $n$. The literature often turns to asymptotic distributions for univariate data \citep{MR1439496}. Here, we propose a \textit{randomized} p-value inspired by conformal prediction methodologies \citep[e.g.][]{vovk2003testing}:
\begin{align*}
p_B=\frac{\sum_{b=1}^{B}\mathbbm{1}\{\a(\pi_b)>\A_n\}+U\big[1+\sum_{b=1}^{B}\mathbbm{1}\{\a(\pi_b)=\A_n\}\big]}{B+1},
\end{align*}
where $\pi_1,\ldots,\pi_B$ are iid from $\mathcal{U}(\Pi_n)$, and $U\sim\mathcal{U}(0,1)$ serves to break ties.

\begin{theorem}\label{thm:H0}
Under $H_0$, with symmetric transformations: (i) $\A_n\overset{d}{=}\a(\pi)$, where $\pi\sim\mathcal{U}(\Pi_n)$; (ii) $p_B\sim\mathcal{U}(0,1)$ for any $B>0$.
\end{theorem}

Theorem \ref{thm:H0} confirms the \text{exact} validity of the p-value $p_B$. Given a prespecified nominal level $\alpha\in(0,1)$, we reject $H_0$ if $p_B<\alpha$. The validity of this p-value is exactly upheld regardless of the parameter $B$. In practice, we recommend a default value of $B=200$.

\renewcommand{\theexample}{\ref{ex:reg}}
\begin{example}[Revisited; High-dimensional setting]
Testing for a change in regression coefficients when $d\gg n$ presents significant challenges. Recently, \cite{zhao2024optimal} introduced a bias-corrected quadratic-form-based CUSUM (QF-CUSUM) test that uses separate LASSO fits at each candidate changepoint, along with a randomization strategy to maintain non-degeneracy under the null hypothesis. \art\ provides simple alternatives: (i) a deviance transformation (with a single global LASSO fit $\widehat{\theta}_{\mathcal{D}}$, see Table \ref{tab:smry}), followed by a nonparametric likelihood aggregation; (ii) a standard $K$-means clustering transformation (see Algorithm \ref{suppsec:clustering} in the supplementary material) applied to the data after discarding all inactive features (those with zero components of $\widehat{\theta}_{\mathcal{D}}$), followed by a rank CUSUM aggregation. Both implementations guarantee exact size control and demonstrate competitive power (see Section \ref{subsec:simu}).
\end{example}

\subsection{Multi-scale adaptations}\label{subsec:multi-scale}

We expands the ideas of ranking and aggregation into a multi-scale framework, laying the groundwork for discussions on multiple changepoint testing, localization, and post-detection inference in Section \ref{sec:MCP}. This framework involves a sequence of $L$ prespecified intervals $\{\mathcal{I}_\ell\subset(0,n]:\ell\in[L]\}$, which may overlap.

Given the transformed scores $\{\S_i:i\in[n]\}$, the \textit{local} rank of $\S_i$ within a specific interval $\mathcal{I}_\ell$, for $i\in\mathcal{I}_\ell$ and $\ell\in[L]$, is defined as:
\[
R_{i,\ell}\equiv R_{i,\mathcal{I}_\ell}=|\{j\in\mathcal{I}_\ell:\S_j\le\S_i\}|.
\]
At each interval, these local ranks are aggregated to construct the statistic $\A_{n,\ell}=\a(\{R_{i,\ell}:i\in\mathcal{I}_\ell\})$. This \textit{multi-scale} local aggregations result in a sequence of statistics $\{\A_{n,\ell}:\ell\in[L]\}$.

According to Theorem \ref{thm:H0}, each statistic $\A_{n,\ell}$ exhibits a marginally distribution-freeness property under $H_0$. However, analyzing the joint distribution of $(\A_{n,1},\A_{n,2},\ldots,\A_{n,L})$ under $H_0$ presents significant challenges due to potential complex dependency among scores and overlapped intervals.

\begin{theorem}\label{thm:multi-scale}
With symmetric transformations and data-independent intervals $\{\mathcal{I}_\ell\}_{\ell=1}^L$:
\begin{itemize}
\item[(i)] (Joint distribution-freeness) Under $H_0$, $(\A_{n,1},\A_{n,2},\ldots,\A_{n,L})\overset{d}{=}\g(\pi)$ for a function $\g:[n]\rightarrow\mathbb{R}^L$, where $\pi\sim\mathcal{U}(\Pi_n)$, and thus its joint distribution is independent of the underlying distribution $P^*_1$.
\item[(ii)] (Pivotalness to changes) If no changepoints are present within any interval $\mathcal{I}_\ell$, the joint distribution of $(\A_{n,1},\A_{n,2},\ldots,\A_{n,L})$ is invariant under both $H_0$ and $H_1$.
\end{itemize}
\end{theorem}

Theorem \ref{thm:multi-scale} is based on a critical observation: The relational ordering between any two scores is directly mirrored in their respective ranks within any given interval. Consequently, we have $R_{i,\ell}=|\{j\in\mathcal{I}_\ell:R_j\le R_i\}|$, which leads to the property of joint distribution-freeness under $H_0$. Similarly, if $\mathcal{I}_\ell\subset(\tau^*_{k-1},\tau^*_k]$ for some $k\in[K^*+1]$, then $R_{i,\ell}=|\{j\in\mathcal{I}_\ell:R_{j,(\tau^*_{k-1},\tau^*_k]}\le R_{i,(\tau^*_{k-1},\tau^*_k]}\}|$. This supports the pivotalness property, further reinforced by the independence of $\{R_{i,(\tau^*_{k-1},\tau^*_k]}:i\in(\tau^*_{k-1},\tau^*_k]\}$ across all true segments $k$. Further details and proofs can be found in the supplementary material.

Building on Theorem \ref{thm:multi-scale}(i), under $H_0$, $\max_{\ell\in[L]}\A_{n,\ell}\overset{d}{=}\|\g(\pi)\|_\infty$. For any $\alpha\in(0,1)$, define:
\begin{align}\label{eq:thresh}
\thresh_{\alpha,B}=\text{the $\lceil(1-\alpha)(B+1)\rceil$-st smallest value among $\{\|\g(\pi_b)\|_\infty:b\in[B]\}$},
\end{align}
where $\pi_1,\ldots,\pi_B$ are iid from $\mathcal{U}(\Pi_n)$. 
\begin{corollary}\label{coro:max_null}
With symmetric transformations and data-independent intervals $\{\mathcal{I}_\ell\}_{\ell=1}^L$, under $H_0$, $\Pr\left\{\max_{\ell\in[L]}\A_{n,\ell}>\thresh_{\alpha,B}\right\} \le\alpha$ for any $\alpha\in(0,1)$ and $B>0$.
\end{corollary}

\section{Multiple changepoint analysis}\label{sec:MCP}

This section builds upon the multi-scale principles established in Theorem \ref{thm:multi-scale} to broaden the scope of the proposed \art\ method across a series of tasks within multiple changepoint analysis. These tasks include multiple changepoint testing, localization with inference, and post-detection inference, while maintaining control over specific global error rates. The enhancements {preserve the model-agnostic and distribution-free attributes of the \art\ test}, ensuring finite-sample guarantees.

\subsection{Multiple changepoint testing}\label{subsec:test}

As discussed in Section \ref{subsec:rank_and_agg}, the \art\ test employs the statistic $\A_n=\a(R_1,\ldots,R_n)$, through aggregations typically using a binary search to identify the most significant locational or distributional changes. This framework is well-suited for scenarios involving at most one change (AMOC). However, its performance tends to diminish in the presence of multiple changepoints, reflecting similar limitations observed with binary segmentation estimation procedures in these contexts \citep{fryzlewicz2014wild}. To address this issue we propose a multi-scale \art\ test specifically configured for assessing the presence of multiple changepoints.

Utilizing the multi-scale framework outlined in Section \ref{subsec:multi-scale}, we engage a sequence of prespecified intervals $\{\mathcal{I}_\ell:\ell\in[L]\}$. From these intervals, we construct a sequence of statistics $\{\A_{n,\ell}:\ell\in[L]\}$ and use the maximum of these statistics 
\[
\A_{n,\rm multi}=\max_{\ell\in[L]}\A_{n,\ell}
\]
as our test statistic. The approach aims to include segments that, by chance, revert to the AMOC scenario where the statistic $\A_{n,\ell}$ proves effective. The selection of intervals may employ moving windows \citep{niu2012screening,eichinger2018mosum}, scanning techniques \citep{MR3076173}, random selection from $(0,n]$ \citep{fryzlewicz2014wild}, or through deterministic strategies within $(0,n]$ \citep{kovacs2023seeded}.

Given $\A_{n,\rm multi}\overset{d}{=}\|\g(\pi)\|_\infty$, where $\pi\sim\mathcal{U}(\Pi_n)$, the associated p-value is readily accessible:
\[
p_{B,\rm multi}=\frac{\sum_{b=1}^{B}\mathbbm{1}\{\|\g(\pi_b)\|_\infty>\A_{n,\rm multi}\}+U\big[1+\sum_{b=1}^{B}\mathbbm{1}\{\|\g(\pi_b)\|_\infty=\A_{n,\rm multi}\}\big]}{B+1},
\]
where $\pi_1,\ldots,\pi_B$ are iid from $\mathcal{U}(\Pi_n)$ and $U\sim\mathcal{U}(0,1)$. We reject the null hypothesis $H_0$ when $p_{B,\rm multi}<\alpha$, given a prespecified nominal Type-I error $\alpha\in(0,1)$.
\begin{theorem}\label{thm:test}
With symmetric transformations and data-independent intervals $\{\mathcal{I}_\ell\}_{\ell=1}^L$, under $H_0$, $\Pr\left\{p_{B,\rm multi}<\alpha\right\}=\alpha$ for any $\alpha\in(0,1)$ and $B>0$. 
\end{theorem}

\subsection{Multiple changepoint localization with inference}\label{subsec:est}

Once $H_0$ is rejected, the subsequent task is to localize multiple changepoints. Recently, Narrowest Significance Pursuit \citep[NSP;][]{fryzlewicz2024narrowest} has been introduced to identify localized regions exhibiting change while maintaining global significance control; see also \cite{fang2020segmentation}. However, these methods often assume specific models (e.g., univariate mean shifts or changes in linear regression coefficients) and rely on known error distributions, limiting their applicability in more general contexts.

To overcome these limitations, we integrate the \art\ methodology with the NSP paradigm. This integration preserves the model-agnostic and distribution-free nature of \art, while maintaining finite-sample control over global false positive rates. The detailed algorithm is presented in Algorithm \ref{alg:est}.

\begin{algorithm}[!ht]
\caption{\art\ localization algorithm.}
\label{alg:est}
\KwInput{Observations $\mathcal{D}=\{Z_i:i\in[n]\}$; symmetric transformation $\s$; rank aggregation $\a$; data-independent intervals $\{\mathcal{I}_\ell:\ell\in[L]\}$; and any algorithm \texttt{SCP} to locate a single changepoint.}

{\textit{Transformation}}: Compute scores $\S_i=\s(Z_i;\mathcal{D})$ for $i\in[n]$.

{\textit{Multi-scale ranking and aggregation}}: Construct statistics $\A_{n,\ell}=\a(\{R_{i,\ell}:i\in\mathcal{I}_\ell\})$ for $\ell\in[L]$, as described in Section \ref{subsec:multi-scale}.

{\textit{Localization}}: Initialize $\widehat{\mathcal{R}}=\emptyset$ and $\widehat{\mathcal{T}}=\emptyset$. Execute $\texttt{NOT}(1,n,\thresh_{\alpha,B})$ with threshold $\thresh_{\alpha,B}$ defined in Eq. (\ref{eq:thresh}).

\Fn{\texttt{NOT}($s,e,\thresh_{\alpha,B}$)}{
\textbf{if} $e-s\leq 1$, \textbf{then} STOP

\Else{
$\mathcal{L}_{s,e}=\{\ell:\mathcal{I}_\ell\subset[s,e]\}$

\textbf{if} $\mathcal{L}_{s,e}=\emptyset$, \textbf{then} STOP

\Else{
$\mathcal{L}_{s,e}^{+}:=\left\{\ell\in\mathcal{L}_{s,e}:\A_{n,\ell}>\thresh_{\alpha,B}\right\}$

\textbf{if} $\mathcal{L}_{s,e}^{+}=\emptyset$, \textbf{then} STOP

\Else{
$\widehat{\ell}=\arg\min_{\ell\in\mathcal{L}_{s,e}^{+}}|\mathcal{I}_\ell|$; denote $\mathcal{I}_{\widehat{\ell}}=[s_{\widehat{\ell}},e_{\widehat{\ell}}]$

Estimate a changepoint $\widehat{\tau}$ within $\mathcal{I}_{\widehat{\ell}}$ via \texttt{SCP} (see Section \ref{subsubsec:consistency})

Update $\widehat{\mathcal{R}}=\widehat{\mathcal{R}}\cup\{\mathcal{I}_{\widehat{\ell}}\}$ and $\widehat{\mathcal{T}}=\widehat{\mathcal{T}}\cup\{\widehat{\tau}\}$

\texttt{NOT}($s,s_{\widehat{\ell}},\thresh_{\alpha,B}$)

\texttt{NOT}($e_{\widehat{\ell}},e,\thresh_{\alpha,B}$)
}
}
}
}

\KwOutput{Localized regions $\widehat{\mathcal{R}}$ and estimated changepoints $\widehat{\mathcal{T}}$.}
\end{algorithm}

This algorithm employs a recursive function $\texttt{NOT}(s,e,\thresh_{\alpha,B})$ (see Line 4), inspired by the Narrowest-Over-Threshold \cite[NOT;][]{baranowski2019narrowest} procedure, to identify the narrowest sub-interval $\mathcal{I}_{\widehat{\ell}}$ for which the statistic $T_{n,\widehat{\ell}}$ exceeds a prespecified threshold $\thresh_{\alpha,B}$ (see Line 10). Although NOT provides large-sample consistency for changepoint estimation with appropriate thresholds, it does not quantify estimation uncertainty. NSP complements NOT for a broad class of linear models by identifying a set of regions, with high probability, each containing at least one true changepoint, thus controlling the global false positive rate. NSP achieves this via a multisolution sup-norm loss-based statistic that measures deviations from linearity, with the threshold determined through simulation assuming a known error distribution. In contrast, \art\ localization is model-agnostic, avoiding linearity assumptions. It remains distribution-free, determining the threshold independently of the data. In practice, the intervals used in the \art\ localization algorithm can be chosen deterministically (e.g., all sub-intervals of $(0,n]$, a sparse set of dyadic intervals \citep{fryzlewicz2024narrowest}, or seeded intervals \citep{kovacs2023seeded}), or selected randomly \citep{fryzlewicz2014wild,baranowski2019narrowest}, offering flexibility in different settings.

\begin{theorem}\label{thm:est}
With symmetric transformations and data-independent intervals $\{\mathcal{I}_\ell\}_{\ell=1}^L$, for any $\alpha\in(0,1)$, $\Pr\big\{\text{there exists $\widehat{R}\in\widehat{\mathcal{R}}$ such that }\widehat{R}\cap\mathcal{T}^*=\emptyset\big\}\le\alpha$, where $\mathcal{T}^*=\{\tau^*_1,\ldots,\tau^*_{K^*}\}$.
\end{theorem}

Theorem \ref{thm:est} establishes that the \art\ localization achieves finite-sample control of the global false positive rate, extending the principle behind NSP into a model-agnostic and distribution-free context. The proof can be outlined as follows:
\begin{align*}
\Pr\big\{&\text{there exists $\widehat{R}\in\widehat{\mathcal{R}}$ such that }\widehat{R}\cap\mathcal{T}^*=\emptyset\big\}\\
&\overset{(i)}{\le}\Pr\Big\{\max_{\ell\in[L];\mathcal{I}_\ell\cap\mathcal{T}^*=\emptyset}\A_{n,\ell}>\thresh_{\alpha,B}\Big\}\\
&\overset{(ii)}{=}\Pr\nolimits_{H_0}\Big\{\max_{\ell\in[L];\mathcal{I}_\ell\cap\mathcal{T}^*=\emptyset}\A_{n,\ell}>\thresh_{\alpha,B}\Big\}
\overset{(iii)}{\le}\Pr\nolimits_{H_0}\Big\{\max_{\ell\in[L]}\A_{n,\ell}>\thresh_{\alpha,B}\Big\}\overset{(iv)}{\le}\alpha.
\end{align*}
Inequality (i) leverages the construction of the \texttt{NOT}-based localization algorithm. Equality (ii) applies the pivotalness property from Theorem \ref{thm:multi-scale}(ii), transferring from $\Pr$ to $\Pr_{H_0}$, thus ensuring a model-agnostic perspective. Inequality (iii) relaxes the subset of intervals to all intervals. Finally, Inequality (iv) follows from Corollary \ref{coro:max_null}, which ensures distribution-free thresholding. In contrast to NSP, where global false positive rate control relies on linearity assumptions and known error distributions, the pivotalness in equality (ii) and the distribution-free threshold determination in inequality (iv) enable \art\ to achieve finite-sample control without imposing such restrictions.

\subsubsection{Changepoint estimation consistency}\label{subsubsec:consistency}

The \art\ localization algorithm produces a set of estimated changepoints from the narrowest intervals $\mathcal{I}_{\widehat{\ell}}$ (see Line 14). The estimation algorithm \texttt{SCP} used can be flexibly chosen, such as simply returning the midpoint of each interval \citep{fryzlewicz2024narrowest}, or in a model-assisted way. To keep the \art\ framework self-contained, we employ an approach based on the aggregation function that seeks to maximize statistics reflecting locational or distributional deviations. For instance, consider the rank CUSUM aggregation applied to an interval $\mathcal{I}_\ell\equiv[s_\ell,e_\ell]$, and define the estimated changepoint as $\widehat{\tau}_{\ell} = \arg\max_{s_\ell\le t<e_\ell}|\mathcal{I}_\ell|^{-3/2}\left\vert\sum_{i=s_\ell}^t\sum_{j=t+1}^{e_\ell}\mathbbm{1}(\S_j\le\S_i)-1/2\right\vert$.

We investigate the consistency of the estimated changepoints through the rates at which the lengths of these intervals contract. For clarity, we focus on the deviance transformation $\s(z;\mathcal{D})=\dev(z;\widehat{f}_\mathcal{D})$, where $\widehat{f}_\mathcal{D}$ is a baseline model that is either prespecified or estimated from the entire data over a model space $\mathcal{F}$. Under the multi-scale ranking and aggregation framework, we consider all sub-intervals of $(0,n]$ and apply the rank CUSUM aggregation. For $k\in[K^*]$, with $Z_i\sim P^*_k$ and $Z_j\sim P^*_{k+1}$, define $Q_k(f)=\Pr\{\dev(Z_j;f)\le\dev(Z_i;f)\}-1/2$, which measures the signal of the change between adjacent segments relative to a candidate model $f\in\mathcal{F}$. Let $\|\cdot\|_\mathcal{F}$ denote the norm induced by the model space $\mathcal{F}$.

\begin{assumption}\label{asmp:est-consistency}
(i) (Baseline model) There exists an $f_0\in\mathcal{F}$ and a constant $c_1>0$ such that $\Pr\{\|\widehat{f}_\mathcal{D}-f_0\|_{\mathcal{F}}\leq c_1\sqrt{(\log n)/n}\}=1+o(1)$. (ii) (Rank function class) The class of functions $\{h_f(z_1,z_2):h_f(z_1,z_2)=\mathbbm{1}(\dev(z_2;f)\le\dev(z_1;f))-1/2,f\in\mathcal{F}\}$ is a Vapnik–Chervonenkis (VC) class of finite VC dimension $\nu$. (iii) (Change spacings) For $k\in[K^*+1]$, $\tau^*_k-\tau^*_{k-1}\geq 2(d_k+d_{k-1})$, where $d_k=\lceil c_2(\log n)/\{Q_k(f_0)\}^2\rceil+1$ for some constant $c_2>0$, with $d_0=d_{K^*+1}=0$. (iv) (Change signal smoothness) For each $k\in[K^*]$, $Q_k(f)$ is Lipschitz continuous, i.e., $|Q_k(f_1)-Q_k(f_2)|\leq c_3\|f_1-f_2\|_{\mathcal{F}}$ for all $f_1,f_2\in\mathcal{F}$, for some constant $c_3>0$. 
\end{assumption}

For a prespecified baseline model $\widehat{f}_\mathcal{D}=f_0$, Assumption \ref{asmp:est-consistency}(i) holds trivially. In finite-dimensional settings, this assumption is typically satisfied under mild conditions, especially when $\widehat{f}_\mathcal{D}$ is a global fit and $f_0$ corresponds to a mixture of underlying segment models. Under simple transformations like $\dev(z;f)=\|z-f\|_2^2$ or $\dev(z,f)=-\phi(z-f)$ for a standard multivariate normal density $\phi$, Assumption \ref{asmp:est-consistency}(ii) follows from standard VC theory \citep[Chapter 4,][]{vapnik1998statistical}. Assumption \ref{asmp:est-consistency}(iii) imposes a minimum spacing requirement between adjacent changepoints on the order of $(\log n)/\min\{\{Q_k(f_0)\}^2,\{Q_{k-1}(f_0)\}^2\}$. In the simple case where $S_i=Z_i\in\mathbb{R}$, $Q_k(f)=\Pr\{Z_j\le Z_i\}-1/2$ corresponds to a familiar measure for locational changes \citep{darkhovskh1976nonparametric}. Assumption \ref{asmp:est-consistency}(iv) requires that $Q_k(f)$ vary smoothly as the model $f$ changes.

\begin{proposition}\label{prop:est-consistency}
Given that Assumption \ref{asmp:est-consistency} holds, as $n\to\infty$, with probability at least $1-\alpha+o(1)$, $\widehat{\mathcal{R}}$ contains exactly $K^*$ intervals $[\widehat{s}_1,\widehat{e}_1]<\cdots<[\widehat{s}_{K^*},\widehat{e}_{K^*}]$ such that $\tau^*_k\in[\widehat{s}_k,\widehat{e}_k)$ and $\widehat{e}_k-\widehat{s}_k\le 2d_k$ for all $k\in[K^*]$.
\end{proposition}

Proposition \ref{prop:est-consistency} parallels to the results on NSP-based localization for linear models \citep{fryzlewicz2024narrowest}, yet under a more general, model-agnostic setting. It demonstrates that the \art\ localization correctly identifies the number of changepoints, each interval returned contains exactly one true changepoint, and the interval length scales on the order of $O\big((\log n)/{Q_k(f_0)}^2\big)$. These findings coincide with the optimal (up to a $\log n$ factor) nonparametric convergence rates reported for univariate changepoint detection problems \citep{MR4255296}, despite differences in how signals are characterized.

\subsection{Inference after multiple changepoint localization}\label{subsec:post}

Section \ref{subsec:est} addresses the localization of regions where changepoints may occur, providing finite-sample global false positive rate control under minimal model and distributional assumptions. To identify a specific changepoint within each region, additional conditions are necessary for consistent estimation. This approach---where interval-based confidence assessments precede the actual localization---is sometimes referred to as \textit{post-inference selection} \citep{fryzlewicz2024narrowest}. In contrast, many practical applications invoke \textit{post-selection} or \textit{post-detection inference}, wherein a set of changepoints is already detected by some method, and the goal is to assess their reliably. Much of the existing literature focus on detection consistency but imposes conditions that may fail in practice---particularly for flexible statistical or machine-learning-based methods. Consequently, some detected changepoints could be inaccurate or even spurious, emphasizing the need for rigorous post-detection inference \citep{hyun2018exact,hyun2021post,duy2020computing,jewell2022testing,carrington2025improving}.

We formalize the problem as follows. Let $\widehat{\mathcal{T}}=\{\widehat{\tau}_j:j\in[\widehat{K}]\}$ be a set of detected changepoints, which may from any detection and localization procedures tailored for specific models. The aim is to test the following sequence of hypotheses \citep{jewell2022testing}:
\begin{align}\label{eq:post_multi_test}
H_{0j}:P^*_{\widehat{\tau}_j-h+1}=\cdots=P^*_{\widehat{\tau}_j+h}\text{ versus $H_{1j}$: not $H_{0j}$,\quad for }j\in[\widehat{K}], 
\end{align}
where $h>0$ defines a prespecified window. If $H_{0j}$ holds, indicating the absence of any true changepoint in $(\widehat{\tau}_j-h,\widehat{\tau}_j+h]$, then $\widehat{\tau}_j$ is treated as a true null. Rejecting $H_{0j}$ labels $\widehat{\tau}_j$ as reliable, and we define the set of such reliable changepoints by $\widehat{\mathcal{T}}_R=\{\widehat{\tau}_j\in\widehat{\mathcal{T}}:H_{0j}\text{ is rejected}\}$.

This post-detection inference task is made difficult by the ``double-dipping'' phenomenon: The same data is used both to select the changepoints and to test their significance. While recent research offers selective p-values that condition on the selection process, those techniques often rely on stringent assumptions tied to specific changepoint models (e.g., Example \ref{ex:mean} where $P_{\theta^*_k}=\mathcal{N}(\theta^*_k,1)$) and detection algorithms.

Alternatively, by recognizing the multiplicity in testing $\{H_{0j}:j\in[\widehat{K}]\}$, \cite{jia2024tune} defines a family-wise error rate (FWER):
$$
\text{FWER}\equiv\Pr\big\{\text{there exists some }\widehat{\tau}_j\in\widehat{\mathcal{T}}_R\cap\mathcal{G}\big\},
$$
where $\mathcal{G}=\left\{h\leq\tau\leq n-h:(\tau-h,\tau+h]\cap\mathcal{T}^*=\emptyset\right\}$ is the collection of all feasible true null changepoints. Here, $H_{0j}$ holds if and only if $\widehat{\tau}_j\in\mathcal{G}$. The TUNE (Thresholding Universally and Nullifying change Effect) procedure introduced by \cite{jia2024tune} rejects $H_{0j}$ whenever a specified test statistic exceeds a universal threshold. TUNE is algorithm-agnostic, allowing the use of any changepoint estimation algorithms. It requires constructing localized two-sample statistics and determining an appropriate threshold by approximating the distribution of the maximum of these statistics under $\Pr_{H_0}$.

Interestingly, the \art\ methodology integrates naturally with TUNE. Concretely, for testing each $H_{0j}$, we use the statistic $\A_{n,\ell}$ on an interval $\mathcal{I}_\ell=(\ell-h,\ell+h]$ and then define
\begin{align}\label{eq:TR}
\widehat{\mathcal{T}}_R=\big\{\widehat{\tau}_j \in \widehat{\mathcal{T}}: \A_{n,\widehat{\tau}_j}>\thresh_{\alpha,B}\big\},
\end{align}
where $\thresh_{\alpha,B}$ is the threshold from Eq. \eqref{eq:thresh}.
\begin{theorem}\label{thm:post}
With symmetric transformations, for any $\alpha\in(0,1)$, ${\rm FWER}\le\alpha$.
\end{theorem}

Theorem \ref{thm:post} shows that the \art\ diagnostic achieves finite-sample FWER control, providing a model-agnostic and distribution-free enhancement of TUNE. Moreover, this integration remains algorithm-agnostic, as $\widehat{\mathcal{T}}$ can be derived from any changepoint detection or localization algorithms. Notably, Theorem \ref{thm:post} mirrors Theorem \ref{thm:est} in spirit---both rely on pivotalness and distribution-freeness---despite targeting entirely different objectives. The proof follows directly from:
\begin{align*}
\text{FWER}&=\Pr\big\{\text{there exists some }\widehat{\tau}_j\in\widehat{\mathcal{T}}_R\cap\mathcal{G}\big\},\\
&\overset{(i)}{\le}\Pr\Big\{\max_{\ell\in[L];\widehat{\tau}_j\in\mathcal{G}}\A_{n,\ell}>\thresh_{\alpha,B}\Big\}\\
&\overset{(ii)}{=}\Pr\nolimits_{H_0}\Big\{\max_{\ell\in[L];\widehat{\tau}_j\in\mathcal{G}}\A_{n,\ell}>\thresh_{\alpha,B}\Big\}
\overset{(iii)}{\le}\Pr\nolimits_{H_0}\Big\{\max_{\ell\in[L]}\A_{n,\ell}>\thresh_{\alpha,B}\Big\}\overset{(iv)}{\le}\alpha.
\end{align*}
Here, Setp (i) follows from the FWER definition and the rule given in \eqref{eq:TR}; steps (ii)--(iv) mirror the arguments in Theorem \ref{thm:est}, leveraging the pivotalness property from Theorem \ref{thm:multi-scale}(ii) and distribution-freeness property from Corollary \ref{coro:max_null}.

\section{Numerical studies}\label{sec:numerical}

We conduct numerical studies to evaluate the performance of the proposed \art\ method across several key tasks in changepoint detection: testing, localization with inference, and post-detection inference. Our experiments cover mean changes, shifts in regression coefficients, and more general distributional changes. Table \ref{tab:smry} outlines the main implementation details for \art, highlighting model-assisted selections of transformation and aggregation choices, as well as multi-scale interval construction for each task. We set $B=200$ throughout, examining \art's robustness to other $B$ values in Section \ref{supp:B} of the supplementary material.

\begin{table}[!tb]
\small
\centering
\caption{Transformation, aggregation, and interval construction for \art.\label{tab:smry}}
\begingroup
\setlength{\tabcolsep}{0pt}
\renewcommand{\arraystretch}{0.9}
\begin{threeparttable} 
\begin{tabular}{cccc}
\toprule
Models & Transformation & Aggregation\tnote{$\star$} & Result \\
\midrule
\multirow{2}{*}{$\substack{\text{Mean} \\ (\text{Example} \ref{ex:mean})}$} & $K$-means\tnote{$\dagger$} & Rank CUSUM & Sections \ref{subsubsec:test}, \ref{subsubsec:post} \\
& $\dev(z;\widehat{f}_\mathcal{D})=z$ & Rank CUSUM & Sections \ref{subsubsec:est}, \ref{subsubsec:welllog} (well-log data) \\
\hline
\multirow{2}{*}{$\substack{\text{Regression} \\ (\text{Example} \ref{ex:reg})}$} & $K$-means & Rank CUSUM & Sections \ref{subsubsec:test}, \ref{subsubsec:est} \\
& $\dev(z;\widehat{f}_\mathcal{D})=(y-x^\top\widehat{\theta}_\mathcal{D})^2$\tnote{$\ddagger$} & Nonparametric likelihood & Section \ref{subsubsec:test} \\
\hline
\multirow{2}{*}{$\substack{\text{Distribution}}$} 
& $\dev(z;\widehat{f}_\mathcal{D})=-\phi(z)$ & Nonparametric likelihood & Sections \ref{subsubsec:test}, \ref{subsubsec:est}\\
& Deep embedded clustering & Rank CUSUM & Section \ref{subsubsec:MNIST} (MNIST data) \\
\bottomrule
\end{tabular}
\begin{tablenotes}[flushleft]\footnotesize
\item[$\dagger$] A standard $K$-means clustering approach is effective for low-dimensional data (see Algorithm \ref{suppsec:clustering} in the supplementary material for details on centroid initialization and cluster number selection). In high-dimensional settings, it can be combined with screening methods (invariant to data order). For instance, in mean change scenarios, retain the dimensions corresponding to the top $10\%$ of entries in $|\widehat{\theta}_{\mathcal{D},j}|$, $j\in[d]$, where $\widehat{\theta}_{\mathcal{D}}=n^{-1}\sum_{i=1}^nZ_i$ is the sample mean. In regression models, discard features for which $\widehat{\theta}_{\mathcal{D},j}=0$, where $\widehat{\theta}_{\mathcal{D}}\in\arg\min_{\theta}\big\{(2n)^{-1}\sum_{i=1}^n(y_i-x_i^\top\theta)^2+\lambda_n\sum_{j=1}^d|\theta_j|\big\}$ is the global LASSO estimate with tuning parameter $\lambda_n=2\sqrt{(\log p)/n}$. Subsequently, the standard $K$-means is applied to the remaining dimensions. Alternatively, screening and $K$-means may be performed jointly, as in the deep embedded clustering for high-dimensional data (see Algorithm \ref{alg:deepembedding}).
\item[$\ddagger$] $\widehat{\theta}_{\mathcal{D}}$ is the global LASSO estimate as defined above.
\item[$\star$] Aggregation involves specifying a sequence of intervals $\{\mathcal{I}_\ell:\ell\in[L]\}$. For multiple changepoint testing, we use moving windows $\{(\ell h-h,\ell h+h]\}_{\ell=1}^{\lfloor(n-h)/h\rfloor}$. For localization with inference, we adopt seeded intervals \citep{kovacs2023seeded}. For post-detection inference, we use $\{(\ell-h,\ell+h]\}_{\ell=h}^{n-h}$.
\end{tablenotes}
\end{threeparttable}
\endgroup
\end{table}

\subsection{Synthetic data}\label{subsec:simu}

Let $0_d\in\mathbb{R}^d$ denote the zero vector, omitting the subscript $d$ when context is clear. Let $I$ be the identify matrix. For a matrix $(a_{ij})\in\mathbb{R}^{d\times d}$, $a_{ij}$ is its $(i,j)$-th element. The following models generate synthetic changes:
\begin{itemize}
\item \underline{Mean changes}: Let $\theta^*_1=0$, and define $\theta^*_{k+1}-\theta^*_k=D_{k,s}$ for $k\in[K^*]$, where $D_{k,s}\in\mathbb{R}^d$ has exactly $s$ nonzero entries, each randomly set to $c_\theta$ or $-c_\theta$, with the rest $0$. Here, $s$ indicates the number of components undergoing a change, and $c_\theta$ specifies its magnitude. Noise is generated from $P_\epsilon=P_{\epsilon,1}^d$, where $P_{\epsilon,1}$ is one of the following distributions: (i) Normal, $c_P\cdot\mathcal{N}(0,1)$; (ii) $t(3)/c_P$, a $t$-distribution with $3$ degrees of freedom; or (iii) Log-normal, $c_P\cdot\exp\{\mathcal{N}(0,1)/10\}$. The constant $c_P$ calibrates the signal-to-noise ratio.
\item \underline{Changes in regression coefficients}: Let $\theta^*_1=(0.5,0,0.5,0_{d-3}^\top)^\top$, and define $\theta^*_{k+1}-\theta^*_k = D_{k,s}$ for $k\in[K^*]$ using the same $D_{k,s}$ construction. Covariates follow $P_x=\mathcal{N}(0,(0.3^{|i-j|}))$, with noise from $P_{\epsilon,1}$ as in the mean change model.
\item \underline{Distributional changes}: We consider three patterns, each switching every two segments: (i) Covariance change: $P^*_{2t-1}=\mathcal{N}(0,c_P\cdot I)$ and $P^*_{2t}=\mathcal{N}(0,(0.9^{|i-j|}))$; (ii) Full change: $P^*_{2t-1}=\mathcal{N}(0,I)$ and $P^*_{2t}=\{t(3)\}^d$; and (iii) Partial change: $P^*_{2t-1}=\mathcal{N}(0,I)$ and $P^*_{2t}=\{t(3)\}^s\cdot\{\mathcal{N}(0,1)\}^{d-s}$, where $s=\lfloor0.4d\rfloor$.
\end{itemize}

Each scenario is repeated $1,000$ times. The subsequent subsections describe each task and its performance metrics.

\subsubsection{Changepoint testing}\label{subsubsec:test}

We set the nominal Type-I error to $\alpha=0.1$ and assess empirical size and power.

\underline{Mean changes}: We compare \art\ with two recently proposed high-dimensional mean change tests that aggregate component-wise CUSUM statistics. The first is the double-max-sum (DMS) test of \cite{wang2023computationally}, which relies on asymptotic theory. The second is the multiplier-bootstrap-based test of \cite{liu2020unified} (LZZL).

\begin{table}[!tb]
\small
\centering
\caption{Empirical size and power of the \art, DMS, and LZZL tests for AMOC scenarios in mean change models under various error distributions.\label{tab:mean_test}}
\begingroup
\setlength{\tabcolsep}{5pt} 
\renewcommand{\arraystretch}{0.9} 
\begin{tabular}{cccccccccccccc}
\toprule
& &     \multicolumn{3}{c}{Null} && \multicolumn{3}{c}{Small change} && \multicolumn{3}{c}{Large change} \\ 
\cline {3-5}  \cline {7-9} \cline{11-13} 
Error &($n$,$d$)  &    
\art\ & DMS & LZZL && \art\ & DMS & LZZL && \art\ & DMS & LZZL  \\ \hline
\multirow{6}{*}{$\substack{\text{Normal} \\ (c_P=0.25)}$}
&(100,100) & 0.101 & 0.135 & 0.144 && 0.264 & 0.292 & 0.319 && 0.588 & 0.576 & 0.665    \\
&(200,100) & 0.096 & 0.117 & 0.133 && 0.481 & 0.525 & 0.542 && 0.826 & 0.941 & 0.955  \\
&(300,100) & 0.099 & 0.103 & 0.113 && 0.652 & 0.771 & 0.775 && 0.969 & 0.995 & 0.999    \\
&(100,200) & 0.095 & 0.134 & 0.152 && 0.205 & 0.225 & 0.299 && 0.531 & 0.488 & 0.593  \\
&(200,200) & 0.094 & 0.127 & 0.121 && 0.423 & 0.440 & 0.432 && 0.755 & 0.871 & 0.904   \\
&(300,200) & 0.098 & 0.111 & 0.096 && 0.619 & 0.654 & 0.664 && 0.874 & 0.989 & 0.997  \\
\hline
\multirow{6}{*}{$\substack{t \\ (c_P=3)}$}
&(100,100) & 0.098 & 0.053 & 0.051 && 0.141 & 0.157 & 0.164 && 0.455 & 0.484 & 0.520   \\
&(200,100) & 0.099 & 0.053 & 0.061 && 0.405 & 0.425 & 0.463 && 0.819 & 0.898 & 0.925  \\
&(300,100) & 0.100 & 0.058 & 0.052 && 0.603 & 0.675 & 0.708 && 0.908 & 0.988 & 0.993   \\
&(100,200) & 0.099 & 0.042 & 0.034 && 0.131 & 0.120 & 0.110 && 0.361 & 0.368 & 0.351  \\
&(200,200) & 0.093 & 0.048 & 0.055 && 0.392 & 0.360 & 0.321 && 0.772 & 0.836 & 0.843   \\
&(300,200) & 0.097 & 0.040 & 0.055 && 0.556 & 0.546 & 0.542 && 0.843 & 0.979 & 0.980  \\
\hline
\multirow{6}{*}{$\substack{\text{Log-normal} \\ (c_P=5)}$}
&(100,100) & 0.097 & 0.136 & 0.127 && 0.220 & 0.311 & 0.338 && 0.483 & 0.658 & 0.690 &  \\
&(200,100) & 0.105 & 0.137 & 0.108 && 0.396 & 0.593 & 0.601 && 0.804 & 0.968 & 0.979 \\
&(300,100) & 0.099 & 0.129 & 0.105 && 0.683 & 0.814 & 0.836 && 0.931 & 0.999 & 1.000   \\
&(100,200) & 0.095 & 0.142 & 0.139 && 0.215 & 0.252 & 0.264 && 0.503 & 0.539 & 0.591  \\
&(200,200) & 0.101 & 0.142 & 0.135 && 0.437 & 0.493 & 0.484 && 0.793 & 0.937 & 0.950  \\
&(300,200) & 0.100 & 0.133 & 0.094 && 0.629 & 0.763 & 0.766 && 0.867 & 0.998 & 0.999   \\
\bottomrule
\end{tabular}
\endgroup
\end{table}

Initially, we consider AMOC scenarios where $n\in\{100,200,300\}$ and $d\in\{100,200\}$. Table \ref{tab:mean_test} shows empirical size and power under various error distributions. Under the null ($K^*=0$), \art\ keeps size near the nominal level in all cases, while DMS and LZZL display mild over-rejection in smaller samples. Under the alternative, we place a single changepoint at $\mathcal{T}^*=\{\lfloor0.4n\rfloor\}$, set $s=3$, and consider small ($c_\theta=0.2$) and large ($c_\theta=0.3$) changes. The \art\ test achieves power comparable to DMS and LZZL, especially as $n$ increases.

\begin{figure}[!ht]
\centering
\includegraphics[scale=0.45]{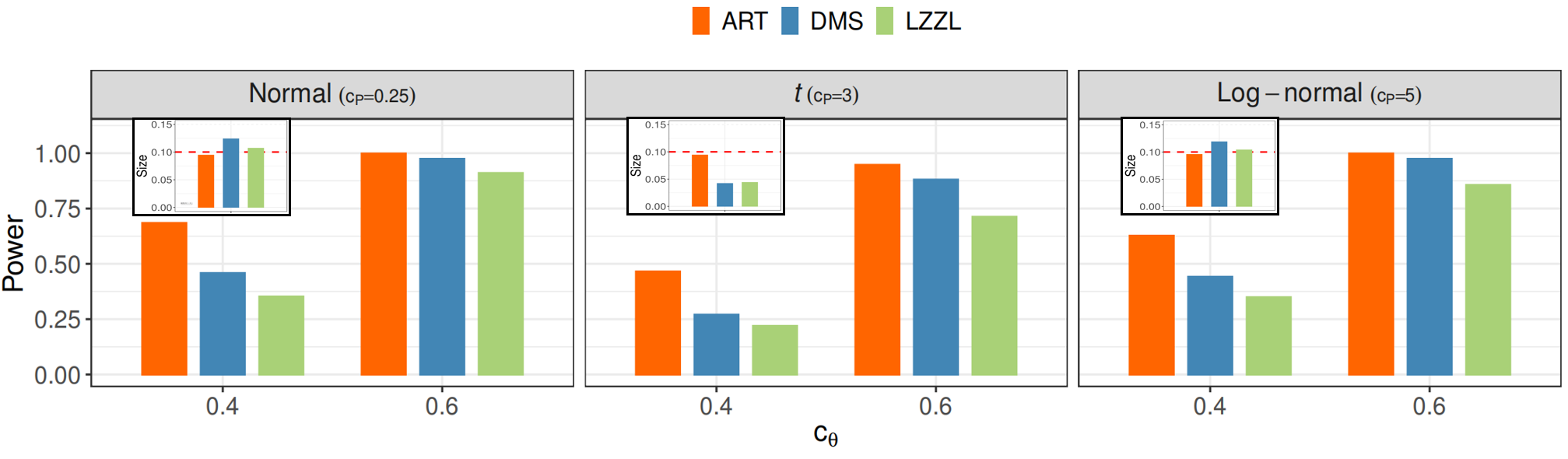}
\caption{Empirical size and power of \art, DMS, and LZZL in mean change models with multiple changepoints under different error distributions. The red dashed line represents the nominal Type-I error.}
\label{fig:mean_test}
\end{figure}

We also examine multiple changepoints for $(n,d)=(200,200)$. Under the alternative, we set $\mathcal{T}^*=\{\lfloor0.3n\rfloor,\lfloor0.4n\rfloor\}$ and $s=3$. For \art, we employ window size $h=0.1n$ (see Table \ref{tab:smry}). Figure \ref{fig:mean_test} illustrates empirical size and power across different error distributions. The \art\ test maintains size close to the nominal level and achieves the highest power, attributable to its multi-scale construction. In contrast, DMS and LZZL use single-scale approaches, which may miss signals if multiple changepoints are present.

\underline{Changes in regression coefficients}: We compare \art\ with the bias-corrected quadratic-form-based CUSUM (QF-CUSUM) test proposed by \cite{zhao2024optimal}, using their recommended tuning parameters.

\begin{table}[!tb]
\small
\centering
\caption{Empirical size and power of the \art\ and QF-CUSUM (QF) tests for AMOC scenarios in regression models under different error distributions. \art\ with clustering and deviance transformations is labeled \art.cl and \art.de, respectively.\label{tab:reg_test}}
\begingroup
\setlength{\tabcolsep}{3pt} 
\renewcommand{\arraystretch}{0.9} 
\begin{tabular}{cccccccccccccc}
\toprule
& &     \multicolumn{3}{c}{Null} && \multicolumn{3}{c}{Small change}  && \multicolumn{3}{c}{Large change} \\ 
\cline {3-5}  \cline {7-9}  \cline{11-13}
Error &$(n, d)$  &    
\art.cl & \art.de & QF  && 
\art.cl & \art.de & QF && 
\art.cl & \art.de & QF \\ \hline
\multirow{6}{*}{$\substack{\text{Normal} \\ (c_P=0.25)}$}
&(100,100) & 0.098 & 0.105 & 0.225 && 0.736 & 0.705 & 0.769 && 0.832 & 0.970 & 0.995 \\
&(200,100) & 0.092 & 0.097 & 0.064 && 0.830 & 0.928 & 0.933  && 0.898 & 0.999 & 1.000  \\
&(400,100) & 0.102 & 0.099 & 0.050   && 0.908 & 0.999 & 0.996  && 0.913 & 1.000 & 1.000 \\
&(100,200) &  0.098 & 0.093& 0.192 && 0.656 & 0.716 & 0.760 && 0.765 & 0.962& 0.990 \\
&(200,200) & 0.095  & 0.098 & 0.056 && 0.791 & 0.916 & 0.799 && 0.877 & 0.997 & 0.999 \\
&(400,200) & 0.096  & 0.105 & 0.033 && 0.893 & 0.997 & 0.955 && 0.899 & 1.000 & 1.000 \\
\hline
\multirow{6}{*}{$\substack{t \\ (c_P=3)}$}
&(100,100) & 0.105 & 0.095 & 0.292 && 0.520 & 0.519 & 0.781 && 0.622 & 0.835 & 0.974  \\
&(200,100) & 0.101 & 0.099 & 0.136 && 0.669 & 0.783 & 0.764 && 0.801 & 0.991 & 0.990    \\
&(400,100) & 0.101 & 0.092 & 0.075 && 0.762 & 0.976 & 0.985 &&  0.877 & 1.000 & 1.000    \\
&(100,200) & 0.100 & 0.101 & 0.299 && 0.423 & 0.509 & 0.724 && 0.580 & 0.818 & 0.960 \\
&(200,200) & 0.087 & 0.102 & 0.095 && 0.625 & 0.773 & 0.715 && 0.795 & 0.985 & 0.973 \\
&(400,200) & 0.115 & 0.099 & 0.073 && 0.759 & 0.971 & 0.935 && 0.874 & 1.000 & 1.000 \\ 
\hline
\multirow{6}{*}{$\substack{\text{Log-normal} \\ (c_P=5)}$}
&(100,100) & 0.093 & 0.099 & 0.083 && 0.494 & 0.449 & 0.317 && 0.615 & 0.825 & 0.903   \\
&(200,100) & 0.098 & 0.102 & 0.007 && 0.630 & 0.714 & 0.393       && 0.826 & 0.989 & 0.989    \\
&(400,100) & 0.098 & 0.102 & 0.002 && 0.754 & 0.940 & 0.546 && 0.887 & 1.000 & 1.000 \\
&(100,200) & 0.095 & 0.094 & 0.083 && 0.418 & 0.466 & 0.246 && 0.588 & 0.820 & 0.877\\
&(200,200) & 0.097 & 0.094 & 0.009 && 0.607 & 0.692 & 0.271 && 0.778 & 0.983 & 0.954\\
&(400,200) & 0.093 & 0.104 & 0.001 && 0.752 & 0.933 & 0.288 && 0.876 & 1.000 & 0.997\\
\bottomrule
\end{tabular}
\endgroup
\end{table}

We study AMOC scenarios with $n\in\{100,200,400\}$ and $d\in\{100,200\}$. Table \ref{tab:reg_test} reports empirical size and power under different error settings. Under the null, \art\ upholds size at the nominal level for all sample sizes, irrespective of the employed transformation, while QF-CUSUM displays size inflation in smaller samples and appears conservative with log-normal errors. Under the alternative, we position a single changepoint at $\mathcal{T}^*=\{\lfloor0.4n\rfloor\}$, set $s=5$, and consider small ($c_{\theta}=0.5$) or large ($c_{\theta}=0.7$) changes. The \art\ test is comparable to QF-CUSUM overall and outperforms it under log-normal errors for small changes. Notably, QF-CUSUM requires ($O(n)$) LASSO fits, bias correction, and sutiable randomization, whereas \art\ with a deviance transformation only needs one global LASSO fit, followed by ranking and aggregation. Despite its simplicity, \art\ maintains valid size and competitive power.

\underline{Distributional changes}: We compare \art\ with two nonparametric changepoint detection methods: ecp (energy-based) \citep{matteson2014nonparametric} and changeforest (random-forest-based) \citep{londschien2023random}. Both ecp and changeforest rely on permutation or pseudo-permutation for significance testing.

We restrict attention to AMOC scenarios with $n\in\{50,100,200\}$ and $d\in\{5,10\}$. Table \ref{tab:dist_test} lists empirical size and power for three types of distributional changes. Under the null, \art\ and ecp maintain size near the nominal level, while changeforest tends to over-reject. Under the alternative (with $\mathcal{T}^*=\{\lfloor0.5n\rfloor\}$), ecp struggles to detect covariance or partial changes, and changeforest attains higher power (but with inflated size). In contrast, \art\ offers stable size control and robust power across the different scenarios.

\begin{table}[!tb]
\small
\centering
\caption{Empirical size and power of the \art, ecp, and changeforest (CF) tests for AMOC scenarios under distributional changes ($c_P=1$).\label{tab:dist_test}}
\begingroup
\setlength{\tabcolsep}{3pt} 
\renewcommand{\arraystretch}{0.9} 
\begin{tabular}{cccccccccccccccccccccccc}
\toprule
&&  \multicolumn{3}{c}{Null} && \multicolumn{3}{c}{Covariance change}  && \multicolumn{3}{c}{Full change} && \multicolumn{3}{c}{Partial change}  \\ 
\cline {3-5}  \cline {7-9}  \cline{11-13} \cline{15-17}
$(n,d)$ &&    
\art\ & ecp & CF  && 
\art\ & ecp & CF && 
\art\ & ecp & CF && 
\art\ & ecp & CF \\ \hline
(50,5)  && 0.104 & 0.110 & 0.157 && 0.442 & 0.262 & 0.917 && 0.905 & 0.535 & 0.982 && 0.343 & 0.112 & 0.802  \\
(100,5) && 0.099 & 0.093 & 0.131 && 0.735 & 0.310 & 1.000 && 1.000 & 0.932 & 1.000 && 0.558 & 0.125 & 0.985 \\
(200,5) && 0.103 & 0.097 & 0.130 && 0.989 & 0.517 & 1.000 && 1.000 & 1.000 & 1.000 && 0.896 & 0.135 & 1.000\\
(50,10)  && 0.097 & 0.101 & 0.149 && 0.455 & 0.320 & 0.990 && 1.000 & 0.572 & 0.999 && 0.526 & 0.113 & 0.947           \\
(100,10) && 0.099 & 0.097 & 0.135 && 0.862 & 0.329 & 1.000 && 1.000 & 0.980 & 1.000 && 0.874 & 0.170 & 1.000 \\
(200,10) && 0.104 & 0.100 & 0.141 && 1.000 & 0.502 & 1.000 && 1.000 & 1.000 & 1.000 && 0.992 & 0.198 & 1.000   \\
\bottomrule
\end{tabular}
\endgroup
\end{table}

\subsubsection{Changepoint localization with inference}\label{subsubsec:est}

\begin{table}[htbp]
\small
\centering
\caption{Comparisons of \art, NSP, and RNSP for changepoint localization with inference across different models and data distributions. The symbol ``-'' indicates scenarios where entries are not applicable.\label{tab:est}}
\begingroup
\setlength{\tabcolsep}{4pt} 
\renewcommand{\arraystretch}{0.85} 
\begin{tabular}{cccccccccc}
\toprule
$\substack{\text{Model} \\ (n,d)}$ & $\substack{Error \\ (c_\theta,c_P)}$ & Method  & FWER & $\text{P}$ & $\text{TP}$ & $\text{TPP}$ & $\text{AveLen}$ & $d_H$ & Time (s) \\ 
\midrule
\multirow{12}{*}{$\substack{\text{Mean} \\ (300,1)}$} & \multirow{3}{*}{$\substack{\text{Normal} \\ (1,1)}$}  & $\text{\art}$ & 0.063 & 1.941 & 1.878 & 0.975 & 42.855 & 15.005 & 3.202 \\
&     & NSP & 0.067 & 1.991 & 1.908 & 0.973 & 46.492 & 20.553 &  1.439 \\
&     & RNSP & 0.002 & 1.664 & 1.662 & 0.999 & 61.525 & 37.157 &  8.841 \\
\cline{3-10}
& \multirow{3}{*}{$\substack{\text{Normal} \\ (2,1)}$}  & $\text{\art}$ & 0.051 & 2.043 & 1.990 & 0.981 & 25.967 &  3.624 & 3.289 \\
&     & $\text{NSP}$ &  0.065 & 2.085 & 1.999 & 0.975 & 15.337 & 7.207 & 1.239 \\
&     & RNSP   & 0.001 & 2.001 & 2.000 & 1.000 & 34.765 & 4.652  & 7.905   \\
\cline{3-10}
& \multirow{3}{*}{$\substack{\text{$t$} \\ (1,\sqrt{3})}$}  & $\text{\art}$ &  0.055 & 2.045 & 1.990 & 0.981 & 34.622 & 5.488 & 3.287  \\
&     & $\text{NSP}$ & 0.993 & 8.291 & 1.712 & 0.240 & 19.978 & 78.799 & 1.126  \\
&     & RNSP   & 0.000 & 1.997 & 1.997 & 1.000 & 47.179 & 8.332 & 7.568 \\
\cline{3-10}
& \multirow{3}{*}{$\substack{\text{$t$} \\ (2,\sqrt{3})}$}  & $\text{\art}$ & 0.060 & 2.058 & 1.996 & 0.979 & 24.418 & 3.878 & 3.271  \\
&     & $\text{NSP}$ & 0.985 & 8.708 & 1.967 & 0.267 & 8.717 & 78.034 & 1.066 \\
&     & RNSP  & 0.002 & 2.002 & 2.000 & 1.000 & 29.993 & 2.544 & 7.375   \\
\hline
\multirow{2}{*}{$\substack{\text{Regression}\\(200,5)}$} & \multirow{12}{*}{$\substack{\text{Normal} \\ (2,1)}$}  & $\text{\art}$ & 0.066 & 1.860 & 1.793 & 0.971 & 37.874 & 14.731 & 3.343  \\
& & $\text{NSP}$ & 0.000 & 1.897 & 1.897 & 1.000 & 36.790 & 15.331 & 26.046 \\
\cline{3-10}
\multirow{2}{*}{$\substack{\text{Regression}\\(400,5)}$} &   & $\text{\art}$ & 0.084 & 2.063 & 1.976 & 0.968 & 44.901 & 8.778 & 6.437   \\
& & $\text{NSP}$ & 0.000 & 1.999 & 1.999 & 1.000 & 40.779 & 12.623 & 35.913   \\
\cline{3-10}
\multirow{2}{*}{$\substack{\text{Regression}\\(200,10)}$} &   & $\text{\art}$ & 0.051 & 1.820 & 1.767 & 0.977 & 39.735 & 16.861 & 4.217   \\
& & $\text{NSP}$ & 0.000 & 1.597 & 1.597 & 1.000 & 55.829 & 29.898 & 38.048 \\
\cline{3-10}
\multirow{2}{*}{$\substack{\text{Regression}\\(400,10)}$}&   & $\text{\art}$ & 0.082 & 2.071 & 1.983 & 0.969 & 46.939 & 8.886 & 7.347   \\
& & $\text{NSP}$ & 0.000 & 1.980 & 1.980 & 1.000 & 65.960 & 18.781 & 48.552 \\
\cline{3-10}
\multirow{2}{*}{$\substack{\text{Regression}\\(200,400)}$}&   & $\text{\art}$ & 0.066 & 1.404 & 1.338 & 0.945 & 51.168 & 39.311 & 6.869   \\
&&  $\text{NSP}$ & - & - & - & - & {-} & {-} & {-}  \\
\cline{3-10}
\multirow{2}{*}{$\substack{\text{Regression}\\(400,400)}$} &   & $\text{\art}$ & 0.068 & 2.029 & 1.958 & 0.972 & 49.920 & 10.538 & 11.171    \\
&&  $\text{NSP}$ & - & - & - & - & {-} & {-} & {-}  \\
\hline
\multirow{6}{*}{$\substack{\text{Distribution}\\(300,30)}$} & \multirow{2}{*}{$\substack{\text{Covariance change} \\ (-,0.25)}$}  
& $\text{\art}$ & 0.054 & 2.057 & 1.999 & 0.981 & 22.437 & 4.114 & 3.581  \\
&     & (R)NSP & - & - & - & - & {-} & {-}  &{-} \\
\cline{3-10}
& \multirow{2}{*}{$\substack{\text{Full change} \\ (-,0.25)}$} 
& $\text{\art}$ & 0.055 & 2.047 & 1.995 & 0.983 & 23.541 & 4.512 & 3.565  \\
&     & (R)NSP   & - & - & - & - & {-} & {-}  &{-} \\
\cline{3-10}
& \multirow{2}{*}{$\substack{\text{Partial change} \\ (-,0.25)}$}  
& $\text{\art}$ & 0.045 & 1.971 & 1.922 & 0.980 & 62.517 & 12.191 & 3.571  \\
&     & (R)NSP   & - & - & - & - & {-} & {-}  &{-}   \\ 
\bottomrule
\end{tabular}
\endgroup
\end{table}

We compare the \art\ localization method with NSP \citep{fryzlewicz2024narrowest} and its robust variant RNSP \citep{fryzlewicz2024robust}, which detects changes in the median (equal to the mean under symmetric errors). Both NSP and RNSP necessitate setting the number of intervals $M$ over which an estimation procedure is iteratively applied. Following \cite{fryzlewicz2024narrowest}, we set $M=1,000$. For \art, we use seeded intervals \citep{kovacs2023seeded} with sets of $\{381,579,778\}$ intervals for $n\in\{200,300,400\}$ respectively. We track several measures:
\begin{itemize}
\item P (positive): number of localized intervals;
\item TP (true positive): number of intervals containing at least one changepoint;
\item TPP (true positive proportion): TP/P;
\item AveLen: average length of intervals deemed true positives;
\item Hausdorff distance: $d_H=\max\left\{\max_{k\in[K^*]}\min_{\widehat{\tau}\in\widehat{\mathcal{T}}}\left|\tau^*_k-\widehat{\tau}\right|,\max_{\widehat{\tau}\in\widehat{\mathcal{T}}}\min_{k\in[K^*]}\left|\tau^*_k-\widehat{\tau}\right|\right\}$, which assesses localization accuracy.
\end{itemize}
Both \art\ and (R)NSP aim to control the probability of localizing intervals that do not contain actual changepoints, thereby controlling ${\rm FWER}=\Pr\{{\rm P}-{\rm TP}>0\}$ at a nominal significance level $\alpha=0.1$.

The underlying changepoint model has two changepoints at $\cT^*=\{\lfloor 0.3n \rfloor, \lfloor0.6n\rfloor \}$ in univariate mean change scenarios, regression coefficient changes ($d\in\{5,10,400\}$, $s=5$), and distributional changes ($d=30$). Note that NSP does not apply to high-dimensional regression or distributional changes, and RNSP is limited to univariate median changes. Table \ref{tab:est} summaries the simulation outcomes for \art, NSP, and RNSP under different models and data distributions. As expected, \art\ controls the FWER below the nominal significance level across all settings, while delivering high TPP, short AveLen, and accurate localization. By contrast:
\begin{itemize}
\item In univariate mean change settings, NSP controls the FWER under normal errors but exhibits severe inflation under $t$-distributed noises. It generally produces shorter intervals (especially for large changes $c_\theta=2$), yet reports more false positives in heavy-tailed settings. RNSP addresses false positives but is more conservative, possibly yielding longer intervals and reduced localization accuracy.
\item For regression coefficient changes under normal errors, NSP slightly surpasses \art\ in TPP and interval lengths when $d$ is small, but intervals widen at higher dimensions.
\end{itemize}
Additionally, \art\ performs only one global score transformation, then computes multi-scale statistics $\A_{n,\ell}$ from local ranks. In contrast, (R)NSP often necessitates repeated multisolution sup-norm loss minimizations \citep{fryzlewicz2024narrowest,fryzlewicz2024robust} over a large collection of intervals, increasing computational burden (see Table \ref{tab:est}), particularly under complex models.

\subsubsection{Post-detection inference}\label{subsubsec:post}

We examine mean change models with $n=600$, $d\in\{10,200\}$, $K^*=3$, and $\mathcal{T}^*=\{\lfloor kn/(K^*+1)\rfloor,k=1,\dots,K^*\}$, setting $s=3$. We apply the Inspect procedure \citep{wang2018high} to detect changepoints, obtaining $\widehat{\cT}$. To validate these detections via the multiple testing framework (\ref{eq:post_multi_test}) (with $h=30$), we use the proposed \art\ diagnostic and TUNE \citep{jia2024tune}. Following \cite{jia2024tune}, TUNE is implemented in two ways: (i) a Wald-type statistic with asymptotic thresholds; and (ii) an $\ell_\infty$-aggregation-based statistic with bootstrap-calibrated thresholds. For \art, we employ a window size $h=30$. Each method is evaluated on two metrics: FWER and power, where
\[
\text{Power}={\bbE}\left[\frac{\left|\left\{\widehat{\tau}_j \in\widehat{\mathcal{T}}: \widehat{\tau}_j\in\widehat{\mathcal{T}}_R,\widehat{\tau}_j\not\in\mathcal{G}\right\}\right|}{\left|\left\{\widehat{\tau}_j\in\widehat{\mathcal{T}}: \widehat{\tau}_j\not\in\mathcal{G}\right\}\right|}\right],
\]
reflecting the expected fraction of detected changepoints deemed reliable. The nominal significance level is set to $\alpha=0.1$.

\begin{figure}[!ht]
\centering
\includegraphics[width=1\textwidth,height=0.48\textwidth]{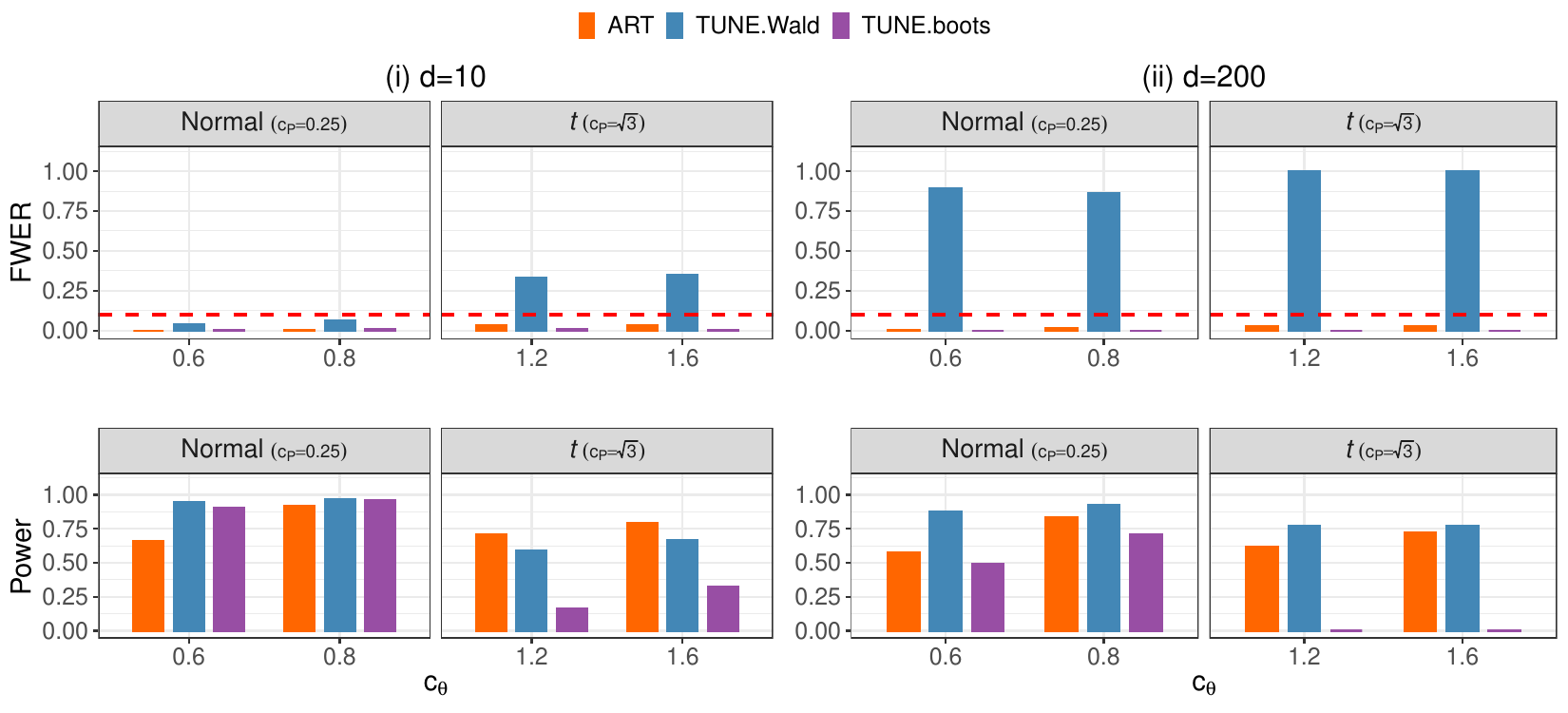}
\caption{Comparisons of \art\ and TUNE (TUNE.Wald for (i) and TUNE.boots for (ii)) for post-detection inference under various error settings. The red dashed line represents the nominal FWER level.}
\label{fig:mean_post}
\end{figure}

Figure \ref{fig:mean_post} compares \art\ and TUNE across different model dimensions and error distributions. In every setting, \art\ keeps the FWER below the nominal level and achieves high power. By contrast, TUNE with Wald thresholds performs well for low-dimensional normal data but inflates FWER in other contexts, suggesting a breakdown of asymptotic approximations. TUNE with bootstrap thresholds preserves valid FWER but is conservative for $t$-distributed errors. These results confirm the distribution-free and model-agnostic benefits of \art\ in post-detection inference.

\subsection{Real-data analysis}\label{subsec:real}

\subsubsection{Well-log dataset}\label{subsubsec:welllog}

We apply the \art\ localization procedure to detect changepoints in a well-log dataset \citep{ruanaidh1996numerical}, which contains $n=4,050$ depth-indexed nuclear magnetic measurements recorded along a borehole. These measurements are commonly assumed to be piecewise constant \citep{fearnhead2019changepoint}, with changepoints indicating geological transitions such as shifts in rock type. Identifying these boundaries informs lithological structure characterization and supports more effective exploration and production decisions.

\begin{figure}[!ht]
\centering
\includegraphics[scale=0.39]{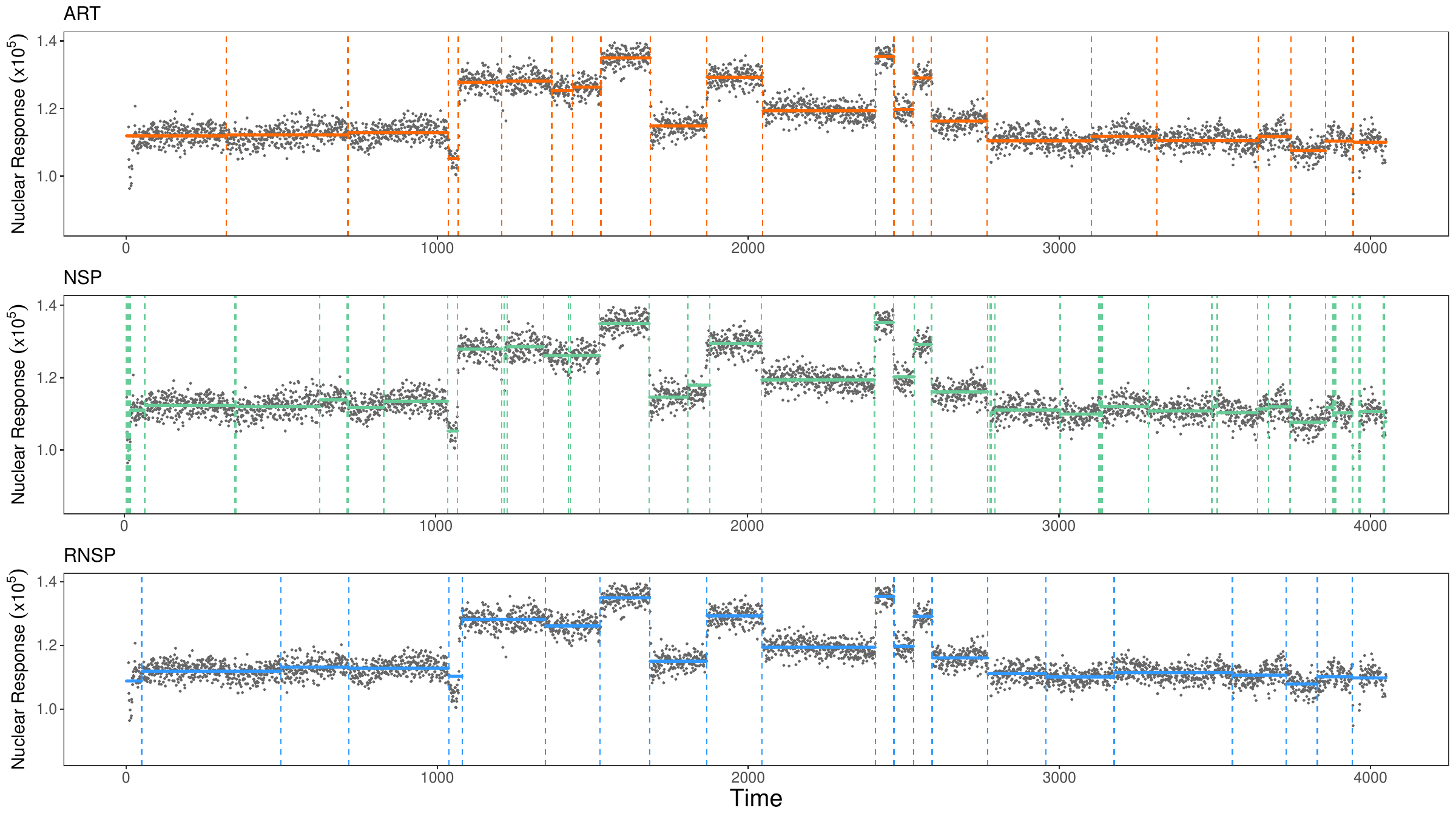}
\caption{The gray dots show the well-log measurements, while the vertical lines mark the detected changepoints at $\alpha=0.1$. The horizontal lines display the median within each segment.}
\label{fig:welllog}
\end{figure}

\art\ follows the guidelines in Table \ref{tab:smry}, using $3,216$ seeded intervals. As benchmarks, we employ NSP and RNSP with $M=3,200$. The nominal significance level is $\alpha=0.1$. \art, NSP, and RNSP detect 23, 43, and 20 changepoints, respectively, with running times (in minutes) 1.106, 5.989, and 2.350. Figure \ref{fig:welllog} presents the data points alongside each method's detected changepoints. NSP appears to overfit because its least-squares-based minimization is sensitive to outliers, producing isolated changepoints. RNSP and \art\ produce similar, more robust detection results.

\subsubsection{MNIST dataset}\label{subsubsec:MNIST}

\begin{figure}[!ht]
\centering
\includegraphics[scale=0.4]{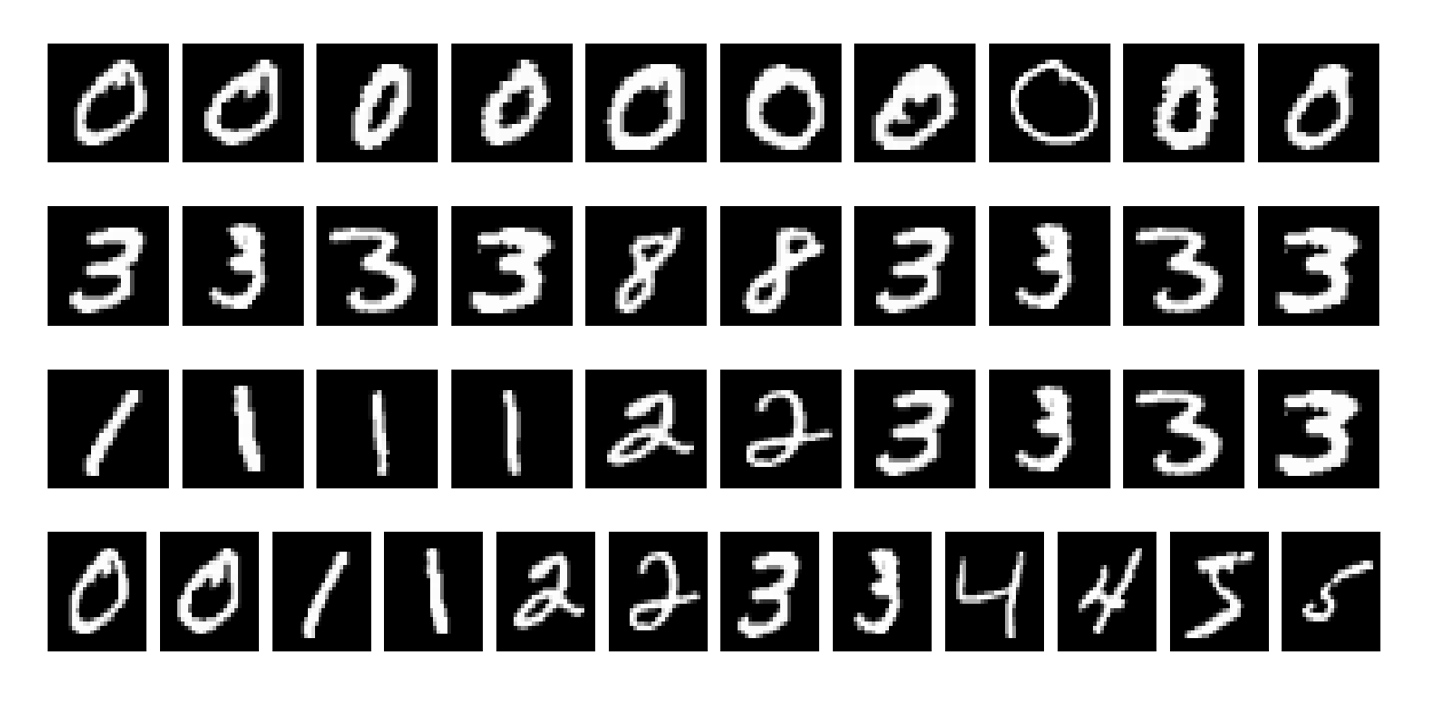}
\caption{Illustration of the four changepoint detection scenarios using the MNIST dataset.}
\label{fig:mnist}
\end{figure}

In this study, we examine the performance of the proposed \art\ method on the high-dimensional, non-Euclidean MNIST dataset \citep{lecun1998gradient}, which comprises 70,000 grayscale images of handwritten digits (0-9), each of size $28\times 28$ pixels. We construct four distinct scenarios (see Figure \ref{fig:mnist}) for changepoint detection in sequences of these images: (i) a control setting with $n$ samples exclusively of digit “0”, containing no changepoints; (ii) a small-signal scenario (3-8-3) where 40\% of the samples are digit “3”, followed by 20\% digit “8”, and then 40\% digit “3” again, reflecting subtle changes due to the visual similarity between “3” and “8”; (iii) a multiple changepoint scenario (1-2-3), comprising 40\% digit “1”, 20\% digit “2”, and 40\% digit “3”; and (iv) a multiple changepoint setting (0-5), with $n$ samples evenly distributed across digits “0” to “5”. In Scenarios (i)--(iii), we set $n=150$; for Scenario (iv), $n=600$. All pixel values are normalized to $[0,1]$. We consider a sequence of tasks---namely, changepoint testing, localization with inference, and post-detection inference---across the four scenarios. \art\ is implemented as described in Table \ref{tab:smry}, employing a transformation based on deep embedded clustering \citep{xie2016unsupervised}, followed by rank CUSUM aggregation. We fix the nominal significance level to $\alpha=0.1$ throughout.

\begin{figure}[!ht]
\centering
\includegraphics[scale=0.6]{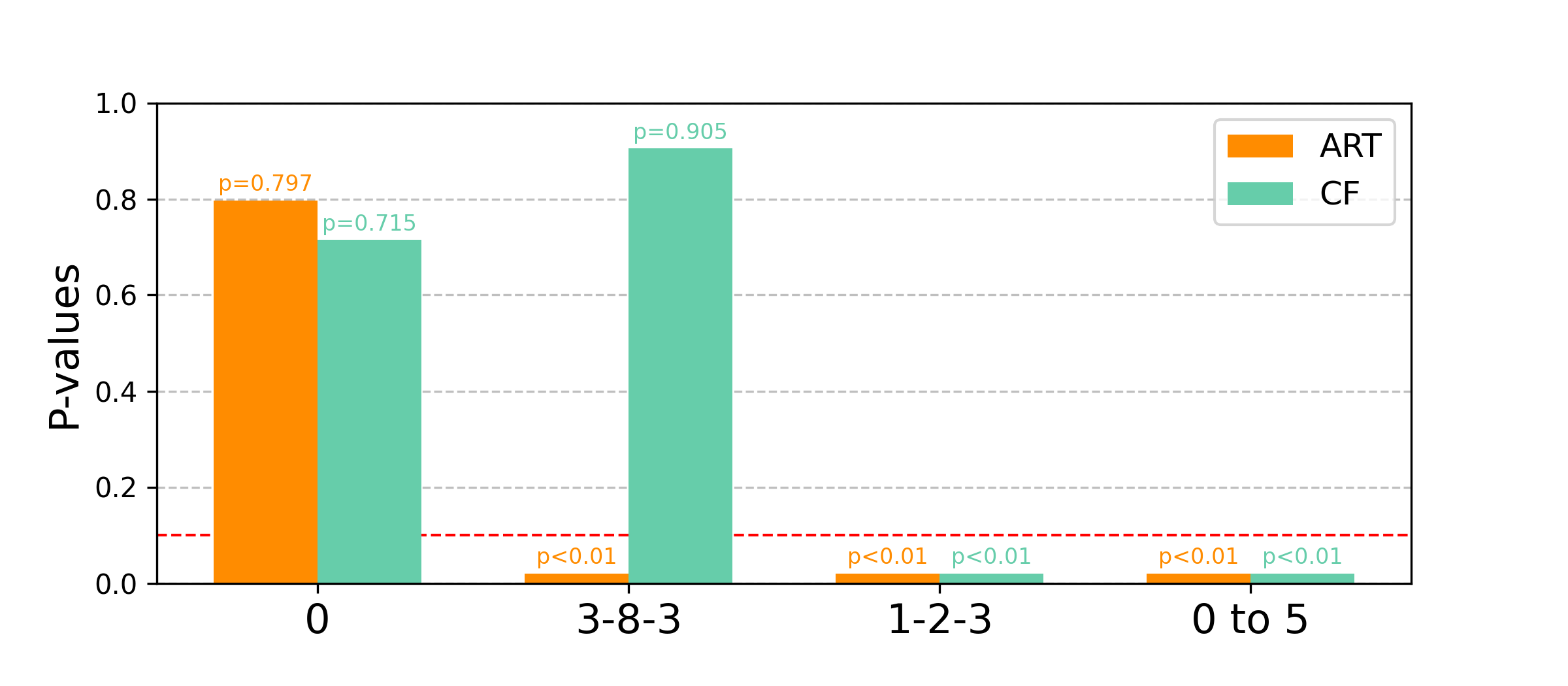}
\caption{Comparison of p-values from the \art\ and changeforest (CF) methods across different MNIST scenarios. The red dashed line indicates the nominal level.}
\label{fig:mnist_test}
\end{figure}

\textbf{Changepoint testing}: We use changeforest \citep{londschien2021change} as a benchmark for changepoint testing; see Section \ref{subsec:test}. For \art, we employ moving windows with $h=30$. Figure \ref{fig:mnist_test} shows that, in the control scenario (0), both approaches yield similarly high p-values, confirming the absence of changepoints. In the small-signal scenario (3-8-3), changeforest fails to detect changepoints (producing a p-value of approximately 0.905), whereas \art\ obtains a p-value below 0.01, successfully signaling changes. In the multiple changepoint scenarios (1-2-3 and 0-5), both methods effectively detect changepoints.

\textbf{Changepoint localization with inference}: After rejecting the null hypothesis in Scenarios (ii)--(iv), we apply the \art\ localization procedure to identify changepoints. We select multi-scale intervals comprising all subintervals of $(0,n]$. Table \ref{tab:mnist_est} summarizes the localization results, demonstrating that \art\ accurately identifies regions containing true changepoints.

\begin{table}[htbp]
\small
\centering
\caption{Localization results for the \art\ method on the MNIST dataset under various changepoint scenarios.\label{tab:mnist_est}}
\begingroup
\setlength{\tabcolsep}{12pt} 
\renewcommand{\arraystretch}{0.9} 
\begin{tabular}{ccccccccccccc}
\toprule
Scenario & $K^*$ & P & TP & TPP & AveLen & $d_H$ \\ \midrule
{3-8-3} & 2 & 2 & 2 & 1 & 32 & 11  \\ 
{1-2-3} & 2 & 2 & 2 & 1 & 18 & 2  \\ 
{0-5} & 5 & 5 & 5 & 1 & 31.2 & 13 \\ 
\bottomrule
\end{tabular}
\endgroup
\end{table}

\textbf{Post-detection inference}: We note that the \art\ localization procedure already provides reliable changepoint estimates, evidenced by the small Hausdorff distance ($d_H$) in Table \ref{tab:mnist_est}. Moreover, \art\ can refine detection results from other methods via post-selection inference. Specifically, we consider the multiple testing framework (\ref{eq:post_multi_test}) with $h=20$, where changepoints are estimated using changeforest and Inspect. The default Inspect implementation often yields redundant, locally clustered changepoints, so we apply a screening procedure: We select the changepoint with the largest CUSUM value, remove any points within $\pm20$ of that location, and repeat until no candidate remains. For \art, we use multi-scale intervals defined by moving windows of size $h$. We track several metrics: P (positive) denotes the number of estimated changepoints or retained changepoints after post-detection inference; TP (true positive) is the number of true changepoints among them; and TPP (true positive proportion) is ${\rm TP}/{\rm P}$. Table \ref{tab:mnist_post} shows that both changeforest and Inspect alone tend to overestimate (leading to large P and $d_H$), despite correctly including all true changepoints (TP = $K^*$). Refining these estimates with \art\ removes false positives, achieving perfect TPP and minimal Hausdorff distance. Hence, \art\ serves as an effective and reliable tool for post-detection inference.

\begin{table}[htbp]
\small
\centering
\caption{Post-detection inference results for the \art\ method compared to changeforest and Inspect on the MNIST dataset.\label{tab:mnist_post}}
\begingroup
\setlength{\tabcolsep}{12pt} 
\renewcommand{\arraystretch}{0.9} 
\begin{tabular}{ccccccccccccc}
\toprule
Scenario & Method & $K^*$ & P & TP & TPP & $d_H$ \\ \hline
\multirow{2}{*}{3-8-3} 
& Inspect & 2 & 4 & 2 & 0.5  & 31  \\ 
& Inspect$+$\art & 2 & 2 & 2 & 1 & 0  \\
\hline
\multirow{4}{*}{1-2-3} 
& CF & 2 & 2 & 2 & 1  & 0  \\
& CF$+$\art & 2 & 2  &  2 & 1 & 0   \\ 
\cline{2-7}
& Inspect & 2 & 4 &  2  & 0.5 & 26    \\ 
& Inspect$+$\art\ & 2 & 2  &  2  &  1 & 0  \\ 
\hline
\multirow{4}{*}{0-5}
& CF & 5 & 8 & 5 & 0.625 & 343   \\
& CF$+$\art & 5 & 5  &  5  & 1  & 0  \\ 
\cline{2-7}
& Inspect & 5 & 17 & 5 & 0.294 & 350   \\ 
& Inspect$+$\art\ & 5 & 5 & 5 & 1 & 1  \\
\bottomrule
\end{tabular}
\endgroup
\end{table}

\section{Conclusion}\label{sec:conclusion}

This paper introduces a distribution-free, model-agnostic approach to changepoint detection that guarantees finite-sample error control. By leveraging symmetric transformations and rank-based aggregations, the proposed framework broadens the applicability of existing methods, offering reliable performance in the absence of stringent distributional or model assumptions. Numerical studies demonstrate that even simple transformations, such as global model fits or naive reference models, can achieve performance comparable to more complex, model-specific approaches. This highlights the practicality of our approach, providing a flexible and robust solution for changepoint detection in complex and less-structured data.

Several avenues for future research remain. First, it would be valuable to explore methods for selecting or constructing optimal or informative score transformations. Additionally, investigating how multiple scores can be efficiently combined to further enhance detection power would be a promising direction. Second, the concept of \textit{pivotalness to changes} provides a novel framework for distribution-free inference in non-exchangeable data settings. This concept may offer new theoretical insights and can potentially be extended to other areas of statistical inference. Finally, extending the proposed framework to online or streaming data scenarios presents a natural and exciting challenge.

\subsection*{Supplementary material}

The supplementary material contains all the theoretical proofs, along with additional algorithmic and numerical details.

\vspace{0.5cm}
{\small \baselineskip 10pt
\bibliographystyle{asa}
\bibliography{ref_change}
}
\newpage
\setcounter{equation}{0}
\def\thelemma{S.\arabic{lemma}}
\def\thepro{S.\arabic{pro}}
\def\theequation{S.\arabic{equation}}
\def\thetable{S.\arabic{table}}
\def\thefigure{S.\arabic{figure}}
\renewcommand{\thealgocf}{S.\arabic{algocf}}
\setcounter{lemma}{0}
\setcounter{figure}{0}
\setcounter{table}{0}
\setcounter{algocf}{0}

\def\thesection{S.\arabic{section}}
\setcounter{section}{0}

\noindent{\bf\Large Supplementary material for ``\art: Distribution-free and model-agnostic changepoint detection with finite-sample guarantees"}

\bigskip

The supplementary material contains all the theoretical proofs, along with additional algorithmic and numerical details. Section \ref{suppsec:clustering} presents implementation details of clustering transformations, including a $K$-means clustering procedure for parametric models (Section \ref{suppsubsec:K-means}) and a deep embedded clustering approach for high-dimensional data (Section \ref{suppsubsec:deepembedding}). Section \ref{suppsec:proofs} collects proofs of Theorems \ref{thm:H0}--\ref{thm:test}, Corollary \ref{coro:max_null}, and Proposition \ref{prop:est-consistency}. Section \ref{suppsec:simulation} shows additional simulation results, including the impact of parameter $B$ in the \art\ method, a comparison between deviance and clustering transformations, and evidence of \art's robustness under departures from data independence.

\section{Clustering transformations}\label{suppsec:clustering}

\subsection{$K$-means clustering for parametric models}\label{suppsubsec:K-means}

Algorithm \ref{alg:K-means} presents a $K$-means clustering-based transformation for constructing scores in parametric models. The algorithm alternates between assigning group memberships and estimating model parameters until convergence.

\begin{algorithm}[ht]
\caption{$K$-means clustering for parametric models.}
\label{alg:K-means}
\KwInput{Observed data $\{Z_i\}_{i=1}^n$; number of clusters $K$; initial cluster centroids $\{f_{j}^{(0)}\}_{j=1}^K$; loss function $\rho(z,f)$; and maximal iteration $M$. Set $k=0$.}

\textbf{Step 1: (Group membership assignment)} For $i\in[n]$, set
\begin{equation*}\label{gi_k}
g_i^{(k+1)}=\underset{j\in[K]}{\arg\min}\ \rho(Z_{i},f_j^{(k)})
\end{equation*}

\textbf{Step 2: (Model parameter estimation)} For $j\in[K]$, set
\begin{equation*}\label{beta_k}
f_j^{(k+1)}=\underset{f\in\mathbb{R}^d}{\arg \min}\sum_{i:g_i^{(k+1)}=j}\rho(Z_{i},f).   
\end{equation*}

\textbf{Step 3:} Set $k=k+1$ and return to Step 1 until group memberships no longer change or the maximal iteration $M$ is reached.

\KwOutput{Scores $\{\s(Z_i;\mathcal{D})=g_i^{(k)}\}_{i=1}^n$.}
\end{algorithm}

\begin{remark}[On the loss function]
Choose loss function $\rho(z,f)$ to suit the specific problem. For instance, in mean estimation, set $\rho(z,f)=\|z-f\|_2^2$, while for regression, use $\rho(z,f)=(y-x^\top f)^2$.
\end{remark}

\begin{remark}[On initial cluster centroids]
For mean models, we use a $K$-means++ initialization \citep{arthur2007k} with a minor modification to ensure a symmetric transformation. Let $D(z)$ be the shortest distance from a point $z$ to the closest centroid already chosen. First, set $f_1^{(0)}=n^{-1}\sum_{i=1}^nZ_i$. Next, select the subsequent centroid as the data point achieving the largest $D(z)$. Repeat until a total of $K$ centroids are obtained.

For regression models with $Z_i=(y_i,x_i)$, we first apply this modified $K$-means++ procedure to $\{\widetilde{z}_i=y_ix_i\}_{i=1}^n$, yielding preliminary centroids. Then perform Step 1 of Algorithm \ref{alg:K-means} with $\rho(\widetilde{z},f)=\|\widetilde{z}-f\|^2_2$ to obtain initial labels. Finally, apply Step 2 of Algorithm \ref{alg:K-means} with $\rho(z,f)=(y-x^\top f)^2$ to update centroids, which serve as the required $\{f_j^{(0)}\}_{j=1}^K$.
\end{remark}

\begin{remark}[On the number of clusters]
When the true number of clusters is unknown, we choose $K$ by minimizing the following Bayesian information criterion: 
\[
\widetilde{K}=\underset{K\in[\bar{K}]}{\arg\min}\bigg\{\frac{n}{2}\log\bigg\{\sum_{j=1}^{K}\sum_{i:g_i^{(k)}(K)=j} \rho(Z_{i},f_j^{(k)}(K))/n\bigg\}+K(d+1)\log n\bigg\},
\]
where $\bar{K}$ is a user-specified upper bound on the number of clusters, and $g_i^{(k)}(K)$ and $f_j^{(k)}(K)$ denote the group memberships and parameters, respectively, from the final iteration of Algorithm \ref{alg:K-means} when run with $K$.
\end{remark}

By construction, the output of Algorithm \ref{alg:K-means} is invariant to the data order, satisfying Definition \ref{def:symmetric_trans}.

\subsection{Deep embedded clustering for high-dimensional data}\label{suppsubsec:deepembedding}

Deep embedded clustering \citep{xie2016unsupervised} is a modern approach that simultaneously learns a lower-dimensional feature representation and a clustering objective. Let $g_\theta:\mathcal{Z}\rightarrow\mathbb{R}^s$ (with $s\ll d$) be a nonlinear map, often implemented through deep neural networks, that transforms the data into a low-dimensional latent feature space. The procedure then iteratively refines both the cluster centroids $\{f_j\}_{j=1}^K$ in the latent space and the parameters $\theta$. Algorithm \ref{alg:deepembedding} outlines this procedure.

\begin{algorithm}[!ht]
\caption{Deep embedded clustering for high-dimensional data.}
\label{alg:deepembedding}
\KwInput{Observed data $\{Z_i\}_{i=1}^n$; and number of clusters $K$.}

\textbf{Step 1: (Parameter initialization)} Initial the nonlinear map parameter $\theta$ and cluster centroids $\{f_j\}_{j=1}^{K}$. See \cite{xie2016unsupervised} for details.

\textbf{Step 2: (Joint optimization)} Minimize the KL divergence:
\[
\text{KL}(P\|Q) = \sum_{i=1}^n \sum_{j=1}^K p_{ij} \log \frac{p_{ij}}{q_{ij}}
\]
via gradient descent, where $q_{ij}$ is a soft assignment (that measures the probability of assigning sample $i$ to cluster $j$) between the embedded points $\{Z'_i=g_\theta(Z_i)\}$ and the cluster centroids $\{f_j\}_{j=1}^{K}$:
\[
q_{ij} = \frac{(1 + \|Z^{\prime}_i - f_j\|_2^2)^{-1}}{\sum_{j'=1}^n (1 + \|Z^{\prime}_i - f_{j'}\|_2^2)^{-1}},
\]
and $p_{ij}$ is an auxiliary distribution:
\[
p_{ij} = \frac{q_{ij}^2 / \sum_{i=1}^n q_{ij}}{\sum_{j'=1}^n \left\{q_{ij'}^2 / \sum_{i=1}^n q_{ij'}\right\}}.
\]

\KwOutput{Scores $\{\s(Z_i;\mathcal{D})=\arg\max_{j\in [K]}q_{ij}\}_{i=1}^n$.}
\end{algorithm}

\section{Proofs}\label{suppsec:proofs}

\subsection{Proof of Theorem \ref{thm:H0}}

\begin{lemma}\label{lemma:Vovk2003}
Let $W_1, W_2, \dots, W_{B+1}$ be iid random variables. Define
\begin{align*}
p_B = \frac{\sum_{b=1}^{B}\mathbbm{1}(W_{B+1}<W_b)+U\left\{1+\sum_{b=1}^{B}\mathbbm{1}(W_{B+1}=W_b)\right\}}{B+1},
\end{align*}
where $U\sim \mathcal{U}(0,1)$ is independent of $\{W_b\}_{b=1}^{B+1}$. For any $\alpha\in (0,1)$, it holds that $\Pr\{p_B<\alpha\}=\alpha$.
\end{lemma}

The details of this lemma's proof appear in the proof of Theorem 1 of \cite{vovk2003testing}. From this lemma, Theorem \ref{thm:H0} follows directly.

\subsection{Proof of Theorem \ref{thm:multi-scale}}

\textbf{Proof of Theorem \ref{thm:multi-scale}(i)}: For any $i\in \mathcal{I}_\ell$, we have $R_{i,\ell}=\sum_{j\in\mathcal{I}_\ell}\mathbbm{1}(\S_j\leq \S_i)=\sum_{j\in\mathcal{I}_\ell}\mathbbm{1}(R_j\leq R_i)$. Then $(\A_{n,1},\A_{n,2},\dots,\A_{n,L})$ can be expressed as a function of $\{R_i\}_{i=1}^n$, say
\[
(\A_{n,1},\A_{n,2},\dots,\A_{n,L})=\g(R_1,\dots,R_n)
\]
for some function $\g:[n]\rightarrow\mathbb{R}^L$. Hence, under $H_0$, $(\A_{n,1},\A_{n,2},\dots,\A_{n,L})\overset{d}{=}\g(\pi)$, where $\pi\sim\mathcal{U}(\Pi_n)$. Therefore, the distribution of $(\A_{n,1},\A_{n,2},\dots,\A_{n,L})$ is independent of $P^*_1$.

\textbf{Proof of Theorem \ref{thm:multi-scale}(ii)}: For notational convenience, denote $R_i^k=R_{i,(\tau^*_{k-1},\tau^*_k]}$ for each $i\in(\tau_{k-1}^*,\tau_{k}^*]$ and $k\in[K^*+1]$.

\begin{lemma}\label{lemma:pivotalness}
The sets of ranks $\{R_{i}^k:i\in(\tau_{k-1}^*,\tau_{k}^*]\}$ are mutually independent across $k\in[K^*+1]$. Moreover, $\widetilde{R}=(R_{1}^1,\dots,R_{\tau^*_1}^1,R_{\tau^*_1+1}^2,\dots,R_{\tau_2^*}^2,\dots,R_{\tau_{K^*}^*+1}^{K^*+1},\dots,R_{\tau_{K^*+1}^*}^{K^*+1})$ has the same distribution under $\Pr$ and $\Pr\nolimits_{H_0}$.
\end{lemma}

Recall that $\A_{n,\ell}=\a(\{R_{i,\ell}\}_{i\in\mathcal{I}_\ell})$. Suppose each interval $\mathcal{I}_\ell$ contains no changepoints. Because $R_{i,\ell}=\sum_{j\in\mathcal{I}_\ell}\mathbbm{1}(\S_j\leq \S_i)=\sum_{j\in\mathcal{I}_\ell}\mathbbm{1}(R_j^k\leq R_i^k)$, $(\A_{n,1},\A_{n,2},\ldots,\A_{n,L})$ can be written as a function of $\widetilde{R}$. By Lemma \ref{lemma:pivotalness}, $(\A_{n,1},\A_{n,2},\dots,\A_{n,L})$ has the same distribution under both $\Pr$ and $\Pr\nolimits_{H_0}$.

\vspace{3mm}
\underline{Proof of Lemma \ref{lemma:pivotalness}:} Without loss of generality, assume there is exactly one changepoint in the data. That is, we demonstrate that $\{R_{i}^1\}_{i\in(0,\tau_{1}^*]}$ and $\{R_{i}^2\}_{i\in(\tau_{1}^*,n]}$ are independent, and that $\widetilde{R}=(R_{1}^1,\dots,R_{\tau^*_1}^1,R_{\tau^*_1+1}^2,\dots,R_{n}^2)$ has the same distribution under $\Pr$ and $\Pr\nolimits_{H_0}$. The extension to multiple changepoints is straightforward.

Denote the row vector $A=(R_{1}^1,R_{2}^1,\dots,R_{\tau^*_1}^1)$ and $B=(R_{\tau^*_1+1}^2,R_{2}^2,\dots,R_{n}^2)$. Let $\pi_1$ be any permutation of $\{1,2,\dots,\tau^*_1\}$ and $\pi_2$ be any permutation of $\{1,2,\dots,n-\tau^*_1\}$. It suffices to show
\begin{align}\label{eq:rank_independence}
\Pr\{A=\pi_1,B=\pi_2\}=\Pr\{A=\pi_1\}\Pr\{B=\pi_2\}.
\end{align}
By the exchangeability of $\S_i$ within each segment, the right-hand side of \eqref{eq:rank_independence} is $\Pr(\{A=\pi_1\}\Pr\{B=\pi_2\}=1/\{\tau^*_1!(n-\tau^*_1)!\}$. For the left-hand side of \eqref{eq:rank_independence}, without loss of generality, choose $\pi_1=(1,2,3,4,\dots,\tau^*_1)$, $\pi_1'=(2,1,3,4,\dots,\tau^*_1)$, and $\pi_2=(1,2,\dots,n-\tau^*_1)$. Then we have
\begin{align*}
&\left\{A=\pi_1,B=\pi_2\right\}
=\left\{\S_1<\S_2<\S_3<\S_4<\dots<\S_{\tau^*_1}\ \text{and} \ \S_{\tau^*_1+1}<\S_{\tau_1+2}<\dots<\S_{n}\right\},\\
&\left\{A=\pi_1',B=\pi_2\right\}=\left\{\S_2<\S_1<\S_3<\S_4<\dots<\S_{\tau^*_1}\ \text{and} \ \S_{\tau^*_1+1}<\S_{\tau_1+2}<\dots<\S_{n}\right\}.
\end{align*}
Because $(Z_1,Z_2,Z_3,Z_4,\dots,Z_n)$ and $(Z_2,Z_1,Z_3,Z_4,\dots,Z_n)$ has the same distribution, Definition \ref{def:symmetric_trans} implies that $(\S_1,\S_2,\S_3,\S_4,\dots,\S_n)$ and $(\S_2,\S_1,\S_3,\S_4,\dots,\S_n)$ also has the same distribution. Consequently, 
\begin{align*}
\Pr\{A=\pi_1,B=\pi_2\}=\Pr\{A=\pi_1',B=\pi_2\}.
\end{align*}
This shows that for any permutation $(\pi_1,\pi_2)$, $\Pr\{A=\pi_1,B=\pi_2\}$ is constant. Moreover, since $\sum_{(\pi_1,\pi_2)}\Pr\{A=\pi_1,B=\pi_2\}=1$ and there are $\tau^*_1!(n-\tau^*_1)!$ possible outcomes for $(\pi_1,\pi_2)$, it follows that $\Pr\{A=\pi_1,B=\pi_2\}=1/\{\tau^*_1!(n-\tau^*_1)!\}$. Hence, \eqref{eq:rank_independence} is established, confirming the independence of $A$ and $B$.

Because $A\sim \mathcal{U}(\Pi_{[\tau^*_1]})$ and $B\sim\mathcal{U}(\Pi_{[n-\tau^*_1]})$ under both $\Pr$ and $\Pr\nolimits_{H_0}$, and $A$ and $B$ are independent, it follows that $\widetilde{R}=(R_{1}^1,\dots,R_{\tau^*_1}^1,R_{\tau^*_1+1}^2,\dots,R_{n}^2)$ has the same distribution under $\Pr$ and $\Pr\nolimits_{H_0}$. 

\subsection{Proof of Corollary \ref{coro:max_null}}

From Theorem \ref{thm:multi-scale}(i), under $H_0$, $\max_{\ell\in[L]}\A_{n,\ell}\overset{d}{=}\|\g(\pi)\|_\infty$. For $\pi_1,\ldots,\pi_B$ iid from $\mathcal{U}(\Pi_n)$ and $\alpha\in(0,1)$,
\begin{align*}
\thresh_{\alpha,B}=\text{the $\lceil(1-\alpha)(B+1)\rceil$-st smallest value among $\{\|\g(\pi_b)\|_\infty:b\in[B]\}$}.
\end{align*}
Then $\max_{\ell\in[L]}\A_{n,\ell}\geq \thresh_{\alpha,B}$ is equivalent to
\begin{align*}
\|\g(\pi)\|_\infty\geq \text{the $\{\lceil(1-\alpha)(B+1) \rceil+1\}$-st smallest value among $\{\|\g(\pi)\|_\infty\}\cup\{\|\g(\pi_b)\|_\infty:b\in[B]\}$}.
\end{align*}
By exchangeability, the rank of $\|\g(\pi)\|_\infty$ in that set is uniform on $\{1,\dots, B + 1\}$. Thus, under $H_0$,
\begin{align*}
\Pr\Big\{\max_{\ell\in[L]}\A_{n,\ell}>\thresh_{\alpha,B}\Big\} \leq \frac{(B+1)-\{\lceil(1-\alpha)(B+1) \rceil+1\}}{B+1}\leq \alpha.
\end{align*}

\subsection{Proof of Theorem \ref{thm:test}}

Under $H_0$, $\A_{n,\rm multi}$ and $\{\|\g(\pi_b)\|_\infty\}_{b=1}^B$ are iid. By Lemma \ref{lemma:Vovk2003}, the result follows immediately.

\subsection{Proof of Proposition \ref{prop:est-consistency}}

Recall that $d_k=\lceil c_2\log n/\{Q_k(f_0)\}^2\rceil+1$ for $k\in[K^*]$, with $d_0=d_{K^*+1}=0$, and $h_f(z_1,z_2)=\mathbbm{1}(\dev(z_2;f)\leq \dev(z_1;f))-1/2$ for $f\in\mathcal{F}$. For $k\in[K^*]$, let $\mathcal{J}_k=(\tau^*_k-d_k,\tau^*_k+d_k]$ and 
\begin{align*}   
S_{\mathcal{J}_k}(f)=\frac{1}{(2d_k)^{3/2}}\sum_{i=\tau^*_k-d_k+1}^{\tau^*_k}\sum_{j=\tau^*_k+1}^{\tau^*_k+d_k}h_f(Z_i,Z_j).
\end{align*}

\begin{lemma}\label{lemma:Sf-ESf}
Under Assumption \ref{asmp:est-consistency}, there exists some constant $c_4>0$ such that
$$
\Pr\Big\{\max_{1\leq k\leq K^*}\sup_{f\in \mathcal{F}}|S_{\mathcal{J}_k}(f)-\mathbb{E}[S_{\mathcal{J}_k}(f)]|\le c_4\sqrt{\log n}\Big\}= 1+o(1).
$$
\end{lemma}

\begin{lemma}\label{lemma:Sf0}
Under Assumption \ref{asmp:est-consistency}, there exist $c_2\geq 8(55+2c_4+c_1c_3)^2$ such that
\begin{align*}
&\Pr\Big\{\min_{1\leq k \leq K^*}\left|S_{\mathcal{J}_k}(f_0)\right| \geq \left(\frac{\sqrt{c_2}}{2\sqrt{2}}-7\right)\sqrt{\log n}\Big\}
\geq 1-6/n^3.
\end{align*}
\end{lemma}

\begin{lemma}\label{lemma:t_alphaB}
Under $H_0$, it holds that 
\begin{align*}
    \Pr\Big\{\thresh_{\alpha,B}<48\sqrt{\log n}\Big\}\geq 1-6B/n^5.
\end{align*}
\end{lemma}

\underline{Proof of Proposition \ref{prop:est-consistency}}: Let $\widehat{\mathcal{R}}=\{\cI_{\ell_k}=[s_{\ell_k},e_{\ell_k}]\}_{k=1}^{\widehat{K}}$. Define
\begin{align*}
&A=\left\{\|\widehat{f}-f_0\|_{\mathcal{F}}\leq c_1\sqrt{(\log n)/n}\right\},\\
&B=\Big\{\max_{1\leq k\leq K^*}\sup_{f\in \mathcal{F}}|S_{\mathcal{J}_k}(f)-\mathbb{E}[S_{\mathcal{J}_k}(f)]|\leq c_4\sqrt{\log n}\Big\},\\
&C=\Big\{\min_{1\leq k \leq K^*}\left|S_{\mathcal{J}_k}(f_0)\right| \geq \left(\frac{\sqrt{c_2}}{2\sqrt{2}}-7\right)\sqrt{\log n}\Big\},\\
&D=\Big\{\thresh_{\alpha,B}<48\sqrt{\log n}\Big\},\\
&E=\Big\{\forall i\in[\widehat{K}],\ \exists k\in[K^*],\text{ such that }\tau^*_k \not\in \cI_{\ell_{i}}\Big\}.
\end{align*}
By Assumption \ref{asmp:est-consistency}(i), Lemmas \ref{lemma:Sf-ESf}--\ref{lemma:t_alphaB}, and Theorem \ref{thm:est}, we have
\begin{align*}
\Pr\left\{A\cap B \cap C \cap D \cap E\right\} \geq 1-\Pr\left\{A^c\cup B^c \cup C^c \cup D^c \cup E^c\right\} \geq 1-\alpha+o(1).
\end{align*}

On $E$, each $\cI_{\ell_{i}}$ contains at least one changepoint. Next, we show each changepoint $\tau_k^*$ is contained in at least one interval $\cI_{\ell_{i}}$. It suffices to verify $\min_{1\leq k \leq K^*}|S_{\mathcal{J}_k}(\widehat{f})|>\thresh_{\alpha,B}$. We have
\begin{align*}
\min_{1\leq k \leq K^*}&|S_{\mathcal{J}_k}(\widehat{f})|\\
&\geq \min_{1\leq k \leq K^*}|S_{\mathcal{J}_k}(f_0)|-\max_{1\leq k \leq K^*}|S_{\mathcal{J}_k}(\widehat{f})-\bbE [S_{\mathcal{J}_k}(\widehat{f})]|-\max_{1\leq k \leq K^*}|\bbE [S_{\mathcal{J}_k}(\widehat{f})]-\bbE [S_{\mathcal{J}_k}(f_0)]|\\
&~~~~~~~~~~~~~~~~~~~~~~~~~ -\max_{1\leq k \leq K^*}|\bbE [S_{\mathcal{J}_k}(f_0)]-S_{\mathcal{J}_k}(f_0)|\\
&\geq \left(\frac{\sqrt{c_2}}{2\sqrt{2}}-7\right)\sqrt{\log n}-c_4\sqrt{\log n}-c_1c_3\sqrt{\log n}-c_4\sqrt{\log n}\\
&> 48\sqrt{\log n},
\end{align*}
using Assumption \ref{asmp:est-consistency}(iv) and the fact that $c_2\geq 8(55+2c_4+c_1c_3)^2$. Event $D$ then implies that $\min_{1\leq k \leq K^*}|S_{\mathcal{J}_k}(\widehat{f})|>\thresh_{\alpha,B}$. Because Algorithm \ref{alg:est} selects the shortest intervals around each exceedance above $\thresh_{\alpha,B}$, each $\tau^*_k$ must lie within an interval of length no more than $|\mathcal{J}_k|=2d_k$. By Assumption \ref{asmp:est-consistency}(iii) and the recursive nature of \art, no interval contains two or more changepoints, and no changepoint is covered by two or more intervals.

Therefore, on $A\cap B \cap C \cap D \cap E$, each $\cI_{\ell_{k}}$ contains exactly one unique changepoint. This establishes a one-to-one correspondence between $\{\cI_{\ell_k}=[s_{\ell_k},e_{\ell_k}]\}_{k=1}^{\widehat{K}}$ and $\{\tau^*_k\}_{k=1}^{K^*}$, with $|e_{\ell_k}-s_{\ell_k}|\leq 2d_k$ for each $k$. This completes the proof.

\subsection{Two-sample U-process theory and proofs of Lemmas \ref{lemma:Sf-ESf}--\ref{lemma:t_alphaB}}

\subsubsection{Two-sample U-process and its properties}

We present definitions and properties of two-sample $U$-statistics and $U$-processes, which will be used to prove Lemmas \ref{lemma:Sf-ESf}--\ref{lemma:t_alphaB}. For a comprehensive introduction, see \cite{clemenccon2021concentration}. Let $X, X_1, \ldots, X_n$ and $Y, Y_1, \ldots, Y_m$ be independent iid samples from distributions $P$ and $Q$ on measurable spaces $\mathcal{X}$ and $\mathcal{Y}$, respectively.

\begin{definition}[Two-sample $U$-statistic/$U$-process of degree $(1,1)$]
Let $h: \mathcal{X} \times \mathcal{Y} \rightarrow \mathbb{R}$ be square-integrable. The two-sample $U$-statistic of degree $(1,1)$ with kernel $h$ is $U_{n, m}(h)=(n m)^{-1} \sum_{i=1}^n \sum_{j=1}^m h\left(X_i, Y_j\right)$. A class of such $U$-statistics indexed by kernels is called a two-sample $U$-process.
\end{definition}

The H\'{a}jek projection of $U_{n, m}(h)-\bbE [U_{n, m}(h)]$ is  
$$
\widehat{U}_{n, m}(h)=(1 / n) \sum_{i=1}^n h_{1,0}\left(X_i\right)+(1 / m) \sum_{j=1}^m h_{0,1}\left(Y_j\right),
$$
where $h_{1,0}(x)=\mathbb{E}_{Y}\left[h\left(x, Y\right)\right]-\mathbb{E}\left[U_{n, m}(h)\right]$ and $h_{0,1}(y)=\mathbb{E}_{X}\left[h\left(X, y\right)\right]-\mathbb{E}\left[U_{n, m}(h)\right]$ for all $(x, y) \in \mathcal{X} \times \mathcal{Y}$. The two-sample $U$-statistic $U_{n, m}(h)$ is called \textit{degenerate} if $h_{1,0}\left(X\right)$ and $h_{0,1}\left(Y\right)$ are almost surely zero. Define $r_{n, m}(h)=U_{n, m}(h)-\mathbb{E}\left[U_{n, m}(h)\right]-\widehat{U}_{n, m}(h)$. Then $r_{n, m}(h)$ is a degenerate two-sample $U$-statistic.

\begin{lemma}[Lemma 27 in \cite{clemenccon2021concentration}]\label{lemma:degenerate_U_tail}
Consider a degenerate two-sample $U$-statistic of degree $(1,1)$ with a bounded kernel $h$ such that $c_{h}=\sup _{(x, y) \in \mathcal{X} \times \mathcal{Y}}|h(x, y)|<\infty$. Then for any $t>0$, $\Pr\left\{U_{n, m}(h) \geq t\right\} \leq \exp\left\{-n m t^2 /\left(32 c_{h}^2\right)\right\}$.
\end{lemma}

\begin{definition}[VC-class]
A collection $\mathcal{F}$ of measurable functions on a sample space is called a VC-class with parameters $a,b>0$ and constant envelope $F>0$ if for any probability measure $Q$,
\[
N\left(\varepsilon F,\mathcal{F},L_2(Q)\right) \leq\left(\frac{a}{\varepsilon}\right)^{b},
\]
for any $\varepsilon \in(0,1)$, where $N\left(\varepsilon,\mathcal{F},L_2(Q)\right)$ defines the covering number.
\end{definition}

By Theorem 2.6.7 in \cite{van1996weak}, a VC-class of dimension $\nu<\infty$ with $F=1$ corresponds to $b=2(\nu-1)$ and $a=\left\{c \nu(16 e)^\nu\right\}^{1 /\{2(\nu-1)\}}$, where $c$ is a universal constant.

\begin{lemma}[Lemma 14 in \cite{clemenccon2021concentration}]\label{lemma:hajek_VC}
Suppose $\mathcal{F}$ is a VC-class of kernels $h: \mathcal{X} \times \mathcal{Y} \rightarrow \mathbb{R}$ with parameters $a, b>0$ and constant envelope. Then, the sets $\{h_{1,0}(x): h \in \mathcal{F}\}$, $\{h_{0,1}(y): h \in \mathcal{F}\}$, and $\{h(x, y)-h_{1,0}(x)-h_{0,1}(y): h \in \mathcal{F}\}$ are also VC-classes with the same parameters $a, b$.
\end{lemma}

\begin{lemma}[Lemma 16 in \cite{clemenccon2021concentration}]\label{lemma:degenerate_Uprocess_tail}
Consider a degenerate two-sample U-process $\left\{U_{n, m}(h): h \in \mathcal{F}\right\}$ of degree $(1,1)$ indexed by asymmetric kernels $h\in\mathcal{F}$ such that $\sup_{(x, y) \in \mathcal{X} \times \mathcal{Y}}|h(x, y)| \leq c_{h}<\infty$ and $\int_{\mathcal{X} \times \mathcal{Y}} h^2(x, y) dP dQ \leq c_{h}^2$. Suppose $\mathcal{F}$ is a VC-class with parameters $a, b>0$ and constant envelope. Then for any $t>0$, there exists a universal constant $C>2$ such that
$$
\Pr\Big\{\sup _{h \in \mathcal{L}}\left|U_{n, m}(h)\right| \geq t\Big\} \leq C 2^{b}(a / c_{h})^{2 b} \exp\left\{4 / c_{h}^2-n m t^2/(w c_{h}^2)\right\},
$$ 
for all $n m t^2>\max \big\{8^4 \log (2) c_{h}^2 b,\left(\log (2) c_{h}^2 b / 2\right)^{1+\zeta}\big\}$ with constants $\zeta \in(1,2)$ and $w=16^3 / 2$.
\end{lemma}

\begin{lemma}[Theorem 2.14.9 in \cite{van1996weak}]\label{lemma:process_tail}
Suppose $\mathcal{F}$ is a VC-class of measurable functions $h: \mathcal{X} \rightarrow [0,1]$ with parameters $a,b>0$ and constant envelope. Then for any $t>0$,
$$
\Pr\Big\{\sup_{h\in\mathcal{F}}\sqrt{n}\Big|\frac{1}{n}\sum_{i=1}^nh(X_i)-\bbE[h(X)]\Big|>t\Big\} \leq\left(\frac{c t}{\sqrt{b}}\right)^b \exp\left\{-2 t^2\right\},
$$
where the constant $c$ depends only on $a$.
\end{lemma}

\subsubsection{Proof of Lemma \ref{lemma:Sf-ESf}}

Recall that 
\begin{align*}
S_{\mathcal{J}_k}(f)
=\frac{1}{(2d_k)^{3/2}}\sum_{i=\tau^*_k-d_k+1}^{\tau^*_k}\sum_{j=\tau^*_k+1}^{\tau^*_k+d_k}h_f(Z_i,Z_j)
=\frac{\sqrt{d_k}}{2\sqrt{2}}U_{d_k, d_k}(h_{f}),
\end{align*}
where $h_{f}(Z_i,Z_j) = \mathbbm{1}(\dev(Z_j;f)\leq \dev(Z_i;f))-1/2$, with $Z_i \sim P_k^*$ and $Z_j \sim P_{k+1}^*$. The H\'{a}jek projection of $U_{d_k, d_k}(h_{f})$ is given by
\[
\widehat{U}_{d_k, d_k}(h_{f})=(1 / d_k) \sum_{i=\tau^*_k-d_k+1}^{\tau^*_k} h_{f,1,0}\left(Z_i\right)+(1 / d_k) \sum_{j=\tau^*_k+1}^{\tau^*_k+d_k} h_{f,0,1}\left(Z_j\right),
\]
where $h_{f,1,0}(z_i)=\mathbb{E}_{Z_{\tau^*_k+1}}\left[h_{f}\left(z_i, Z_{\tau^*_k+1}\right)\right]-\mathbb{E}\left[U_{d_k, d_k}(h_{f})\right]$ and  $h_{f,0,1}(z_j)=\mathbb{E}_{Z_{\tau^*_k}}\left[h_{f}\left(Z_{\tau^*_k}, z_j\right)\right]-\mathbb{E}\left[U_{d_k, d_k}(h_{f})\right]$. By Assumption \ref{asmp:est-consistency}(ii), $\{h_f:f\in\mathcal{F}\}$ is a VC-class with parameters $a=\left\{c \nu(16 e)^\nu\right\}^{1 /\{2(\nu-1)\}}$, $b=2(\nu-1)$, and constant envelope $1$.

Denote $r_{d_k, d_k}(h_{f})=U_{d_k, d_k}(h_{f})-\mathbb{E}\left[U_{d_k, d_k}(h_{f})\right]-\widehat{U}_{d_k, d_k}(h_{f})$, then
\begin{align*}
r_{d_k, d_k}(h_{f})&=\frac{1}{d_k^2}\sum_{i=\tau^*_k-d_k+1}^{\tau^*_k}\sum_{j=\tau^*_k+1}^{\tau^*_k+d_k} h_{f}\left(Z_i, Z_j\right)-\bbE [h_{f}\left(Z_i, Z_j\right)]-h_{f,1,0}\left(Z_i\right)-h_{f,0,1}\left(Z_j\right)\\
&=\frac{1}{d_k^2}\sum_{i=\tau^*_k-d_k+1}^{\tau^*_k}\sum_{j=\tau^*_k+1}^{\tau^*_k+d_k} h_{f}'(Z_i, Z_j),
\end{align*}
where $h_{f}'(Z_i, Z_j)=h_{f}\left(Z_i, Z_j\right)-\bbE [h_{f}\left(Z_i, Z_j\right)]-h_{f,1,0}\left(Z_i\right)-h_{f,0,1}\left(Z_j\right)$. Notice that $\{r_{d_k, d_k}(h_{f}): f\in \mathcal{F}\}$ is a degenerate two-sample $U$-process. By Lemma \ref{lemma:hajek_VC}, $\{h_{f}':f\in \mathcal{F}\}$ is a VC-class with parameter $(a,b)$. Noticing that $|h_{f}'(Z_i, Z_j)|\leq 3$, Lemma \ref{lemma:degenerate_Uprocess_tail} implies that
\begin{align*}
\Pr\bigg\{\sup_{f\in \mathcal{F}}\Big|U_{d_k, d_k}(h_{f})&-\mathbb{E}\left[U_{d_k, d_k}(h_{f})\right]-\widehat{U}_{d_k, d_k}(h_{f})\Big|\geq x\bigg\}\\
&\leq C 2^{b}(a / c_{h})^{2 b} \exp\left\{4 / c_{h}^2-d_k^2 x^2/(w c_{h}^2)\right\},
\end{align*} 
for all $d_k^2 x^2>\max \big\{8^4 \log (2) c_{h}^2 b,\left(\log (2) c_{h}^2 b / 2\right)^{1+\zeta}\big\}$.
Hence,
\begin{align*}
\Pr\bigg\{\sup_{f\in \mathcal{F}}\Big|S_{\mathcal{J}_k}(f)&-\mathbb{E}\left[S_{\mathcal{J}_k}(f)\right]-\frac{\sqrt{d_k}}{2\sqrt{2}}\widehat{U}_{d_k, d_k}(h_{f})\Big| \geq x\bigg\}\\
&\leq C 2^{b}(a / c_{h})^{2 b} \exp\left\{4 / c_{h}^2-8d_k x^2/(w c_{h}^2)\right\},
\end{align*}
for all $8d_k x^2>\max \big\{8^4 \log (2) c_{h}^2 b,\left(\log (2) c_{h}^2 b / 2\right)^{1+\zeta}\big\}$.

Next, consider the H\'{a}jek projection $\widehat{U}_{d_k, d_k}(h_{f})$. Direct calculation shows $|h_{f,1,0}\left(Z_i\right)|\leq 1$, $|h_{f,0,1}\left(Z_j\right)|\leq 1$, $\bbE[h_{f,1,0}\left(Z_i\right)]=0$, and $\bbE[h_{f,0,1}\left(Z_j\right)]=0$. By Lemma \ref{lemma:hajek_VC}, both $\{h_{f,1,0}:f\in \mathcal{F}\}$ and $\{h_{f,0,1}:f\in \mathcal{F}\}$ are VC-classes with parameter $(a,b)$. Applying Lemma \ref{lemma:process_tail}, we have
\begin{align*}
\Pr\bigg\{\sup_{f\in \mathcal{F}}\sqrt{d_k}\bigg|\frac{1}{d_k} \sum_{i=\tau^*_k-d_k+1}^{\tau^*_k} h_{f,1,0}\left(Z_i\right)\bigg| \geq x\bigg\}
&\leq \left(\frac{c x}{\sqrt{b}}\right)^b \exp\left\{-2 x^2\right\},\\
\Pr\bigg\{\sup_{f\in \mathcal{F}}\sqrt{d_k}\bigg|\frac{1}{d_k} \sum_{j=\tau^*_k+1}^{\tau^*_k+d_k} h_{f,0,1}\left(Z_j\right)\bigg| \geq x\bigg\}
&\leq \left(\frac{c x}{\sqrt{b}}\right)^b \exp\left\{-2 x^2\right\}.
\end{align*}
This further implies that
\begin{align*}
\Pr\bigg\{\sup_{f\in \mathcal{F}}\bigg|\frac{\sqrt{d_k}}{2\sqrt{2}}\widehat{U}_{d_k, d_k}(h_{f_0})\bigg| \geq x\bigg\} \leq 2\left(\frac{c x}{\sqrt{b}}\right)^b \exp\left\{-4 x^2\right\}.
\end{align*}

By the triangle inequality,
\begin{align*}
\Pr\bigg\{\max_{1\leq k\leq K^*}&\sup_{f\in \mathcal{F}}\left|S_{\mathcal{J}_k}(f)-\mathbb{E}\left[S_{\mathcal{J}_k}(f)\right]\right|\geq 2x\bigg\}\\
&\leq \Pr\bigg\{\max_{1\leq k\leq K^*}\sup_{f\in \mathcal{F}}\bigg|S_{\mathcal{J}_k}(f)-\mathbb{E}\left[S_{\mathcal{J}_k}(f)\right]-\frac{\sqrt{d_k}}{2\sqrt{2}}\widehat{U}_{d_k, d_k}(h_{f})\bigg|\geq x\bigg\}\\
&~~~~~~~~~~~~~~~~~~~~+\Pr\bigg\{\max_{1\leq k\leq K^*}\sup_{f\in \mathcal{F}}\left|\frac{\sqrt{d_k}}{2\sqrt{2}}\widehat{U}_{d_k, d_k}(h_{f})\right|\geq x\bigg\}\\
&\leq K^*C 2^{b}(a / c_{h})^{2 b} \exp\left\{4 / c_{h}^2-8d_k x^2/(w c_{h}^2)\right\}+2K^*\left({c x}/{\sqrt{b}}\right)^b \exp\left\{-4 x^2\right\}.
\end{align*}
Setting $2x=c_4\sqrt{\log n}$ finishes the argument, showing that
\[
\Pr\left\{\max_{1\leq k\leq K^*}\sup_{f\in \mathcal{F}}\left|S_{\mathcal{J}_k}(f)-\mathbb{E}[S_{\mathcal{J}_k}(f)]\right|< c_4\sqrt{\log n}\right\}=1+o(1)
\]
for some constant $c_4>0$.

\subsubsection{Proof of Lemma \ref{lemma:Sf0}}

Arguing as in the proof of Lemma \ref{lemma:Sf-ESf} but applying Lemma \ref{lemma:degenerate_U_tail}, we get
\begin{align*}
\Pr\left\{\left|U_{d_k, d_k}(h_{f_0})-\mathbb{E}\left[U_{d_k, d_k}(h_{f_0})\right]-\widehat{U}_{d_k, d_k}(h_{f_0})\right| \geq x\right\} \leq 2\exp\left\{-d_k^2 x^2 /72\right\},
\end{align*}
which implies
\begin{align*}
\Pr\bigg\{\bigg|S_{\mathcal{J}_k}(f_0)-\frac{\sqrt{d_k}}{2\sqrt{2}}\mathbb{E}\left[U_{d_k, d_k}(h_{f_0})\right]-\frac{\sqrt{d_k}}{2\sqrt{2}}\widehat{U}_{d_k, d_k}(h_{f_0})\bigg| \geq x\bigg\} \leq 2\exp\left\{-d_k x^2 /9\right\}.
\end{align*}

Consider the H\'{a}jek projection
\[
\widehat{U}_{d_k, d_k}(h_{f_0})=(1 / d_k) \sum_{i=\tau^*_k-d_k+1}^{\tau^*_k} h_{f_0,1,0}\left(Z_i\right)+(1 / d_k) \sum_{j=\tau^*_k+1}^{\tau^*_k+d_k} h_{f_0,0,1}\left(Z_j\right).
\]
By Hoeffding's inequality,
\begin{align*}
\Pr\bigg\{\bigg|\frac{1}{d_k} \sum_{i=\tau^*_k-d_k+1}^{\tau^*_k} h_{f,1,0}\left(Z_i\right)\bigg| \geq x\bigg\}
&\leq 2\exp\left\{-2d_k x^2\right\},\\  
\Pr\bigg\{\bigg|\frac{1}{d_k} \sum_{j=\tau^*_k+1}^{\tau^*_k+d_k} h_{f,0,1}\left(Z_j\right)\bigg| \geq x\bigg\}
&\leq 2\exp\left\{-2d_k x^2\right\}.
\end{align*}
Thus,
\begin{align*}
\Pr\left\{\left|\frac{\sqrt{d_k}}{2\sqrt{2}}\widehat{U}_{d_k, d_k}(h_{f_0})\right| \geq x\right\} \leq 4\exp\left\{-4 x^2\right\}.
\end{align*}

By the triangle inequality,
\begin{align*}
\Pr\Big\{\Big|S_{\mathcal{J}_k}(f_0)\Big| &< \Big|\frac{\sqrt{d_k}}{2\sqrt{2}}\mathbb{E}\Big[U_{d_k, d_k}(h_{f_0})\Big]\Big|-\sqrt{\log n}-\frac{6\sqrt{\log n}}{\sqrt{d_k}}\Big\}\\
&\leq \Pr\Big\{\Big|S_{\mathcal{J}_k}(f_0)-\frac{\sqrt{d_k}}{2\sqrt{2}}\mathbb{E}\Big[U_{d_k, d_k}(h_{f_0})\Big]\Big|\geq \sqrt{\log n}+\frac{6\sqrt{\log n}}{\sqrt{d_k}}\Big\}\\
&\leq \Pr\Big\{\Big|S_{\mathcal{J}_k}(f_0)-\frac{\sqrt{d_k}}{2\sqrt{2}}\mathbb{E}\Big[U_{d_k, d_k}(h_{f_0})\Big]-\frac{\sqrt{d_k}}{2\sqrt{2}}\widehat{U}_{d_k, d_k}(h_{f_0})\Big| \geq \frac{6\sqrt{\log n}}{\sqrt{d_k}}\Big\}\\
&~~~~~~~~~~~~~~~~~~~~~ + \Pr\Big\{\Big|\frac{\sqrt{d_k}}{2\sqrt{2}}\widehat{U}_{d_k, d_k}(h_{f_0})\Big| \geq \sqrt{\log n}\Big\}\\
&\leq 2\exp(-4\log n) + 4\exp(-4\log n)=6/n^4.
\end{align*}
Note that $\mathbb{E}\left[U_{d_k, d_k}(h_{f_0})\right]=Q_k(f_0)$. By definition $d_k=\lceil c_2\log n/\{Q_k(f_0)\}^2\rceil+1$, we have
$$
\left|\frac{\sqrt{d_k}}{2\sqrt{2}}\mathbb{E}\left[U_{d_k, d_k}(h_{f_0})\right]\right|> \frac{\sqrt{c_2}}{2\sqrt{2}}\sqrt{\log n},
$$
implying
\begin{align*}
&\Pr\left\{\left|S_{\mathcal{J}_k}(f_0)\right| \geq \left(\frac{\sqrt{c_2}}{2\sqrt{2}}-7\right)\sqrt{\log n}\right\}
\geq  1-6/n^4.
\end{align*}
A union bound then gives
\begin{align*}
\Pr\Big\{\min_{1\leq k \leq K^*}&\left|S_{\mathcal{J}_k}(f_0)\right| \geq \left(\frac{\sqrt{c_2}}{2\sqrt{2}}-7\right)\sqrt{\log n}\Big\}\\
&\geq 1-\Pr\Big\{\exists k \in [K^*] \text{ such that } \left|S_{\mathcal{J}_k}(f_0)\right| < \left(\frac{\sqrt{c_2}}{2\sqrt{2}}-7\right)\sqrt{\log n}\Big\}\\
&\geq 1-6/n^3.
\end{align*} 

\subsubsection{Proof of Lemma \ref{lemma:t_alphaB}}

Recall that under $H_0$, $\max_{\ell\in[L]}\A_{n,\ell}\overset{d}{=}\|\g(\pi)\|_\infty$, and
\begin{align*}
\thresh_{\alpha,B}=\text{the $\lceil(1-\alpha)(B+1)\rceil$-st smallest value among $\{\|\g(\pi_b)\|_\infty:b\in[B]\}$},
\end{align*}
where for any $\mathcal{I}_\ell=[\sl,\el]$, 
\begin{align*}
\A_{n,\ell}=\sup_{t\in[\sl,\el]}\bigg|\frac{1}{(\el-\sl)^{3/2}}\sum_{i=\sl}^t\sum_{j=t+1}^{\el}\left(\mathbbm{1}(\S_j\leq \S_i)-\frac{1}{2}\right)\bigg|.
\end{align*}
The proof strategy is to characterize the tail probability of $\A_{n,\ell}$. Although $(\S_1,\S_2,\dots,\S_n)$ is exchangeable under $\Pr\nolimits_{H_0}$, $\A_{n,\ell}$ is not a two-sample $U$-statistic. However, treating $\S_1,\S_2,\dots,\S_n$ as iid continuous random variables leaves the distribution of 
$\A_{n,\ell}$ unchanged. Hence, in the following we assume 
$\S_1,\S_2,\dots,\S_n$ are iid continuous random variables.

For notational convenience, define
\begin{align*}
S(\sl,t,\el)&=\frac{1}{(\el-\sl)^{3/2}}\sum_{i=\sl}^t \sum_{j=t+1}^{\el}h(S_i,S_j),\\
U(\sl,t,\el)&=\frac{1}{(t-\sl+1)(\el-t)}\sum_{i=\sl}^t \sum_{j=t+1}^{\el}h(S_i,S_j),
\end{align*}
where $h(S_i,S_j)=\mathbbm{1}(\S_j\leq \S_i)-1/2$.
Because $\mathbb{E}[U(\sl,t,\el)]=0$, applying Lemma \ref{lemma:degenerate_U_tail} yields
\begin{align*}
\Pr\nolimits_{H_0}\left\{\left|U(\sl,t,\el)-\widehat{U}(\sl,t,\el)\right| \geq x\right\} \leq 2\exp\left\{-(t-\sl+1)(\el-t) x^2 /72\right\},
\end{align*}
where $\widehat{U}(\sl,t,\el)=(1 / (t-\sl+1)) \sum_{i=\sl}^t h_{1,0}\left(\S_i\right)+(1 / (\el-t)) \sum_{j=t+1}^{\el} h_{0,1}\left(\S_j\right)$ is the H\'{a}jek projection of $U(\sl,t,\el)$. It follows that
\begin{align*}
\Pr\nolimits_{H_0}\left\{\left|S(\sl,t,\el)-\frac{(t-\sl+1)(\el-t)}{(\el-\sl)^{3/2}}\widehat{U}(\sl,t,\el)\right| \geq x\right\}
\leq  2\exp\left\{- x^2 /72\right\}.
\end{align*}

Consider the H\'{a}jek projection $\widehat{U}(\sl,t,\el)$ and applying Hoeffding's inequality in a manner similar to Lemma \ref{lemma:Sf0}, we obtain
\begin{align*}
\Pr\nolimits_{H_0}\bigg\{\bigg|\frac{1}{t-\sl+1} \sum_{i=\sl}^t h_{1,0}\left(Z_i\right)\bigg| \geq x\bigg\} &\leq 2\exp\left\{-2(t-\sl+1)) x^2\right\}, \\
\Pr\nolimits_{H_0}\bigg\{\bigg|\frac{1}{\el-t} \sum_{j=t+1}^{\el} h_{0,1}\left(Z_j\right)\bigg| \geq x\bigg\} &\leq 2\exp\left\{-2(\el-t) x^2\right\}.
\end{align*}
Hence,
\begin{align*}
\Pr\nolimits_{H_0}\bigg\{\frac{(t-\sl+1)(\el-t)}{(\el-\sl)^{3/2}}\bigg|\frac{1}{t-\sl+1} \sum_{i=\sl}^t h_{1,0}\left(Z_i\right)\bigg| \geq x\bigg\} 
&\leq 2\exp\left\{- x^2\right\},\\
\Pr\nolimits_{H_0}\bigg\{\frac{(t-\sl+1)(\el-t)}{(\el-\sl)^{3/2}}\bigg|\frac{1}{\el-t} \sum_{j=t+1}^{\el} h_{0,1}\left(Z_j\right)\bigg| \geq x\bigg\}
&\leq 2\exp\left\{- x^2\right\}.
\end{align*}
A triangle inequality gives
\begin{align*}
\Pr\nolimits_{H_0}\left\{\left|\frac{(t-\sl+1)(\el-t)}{(\el-\sl)^{3/2}}\widehat{U}(\sl,t,\el)\right| \geq x\right\} \leq 4\exp\left\{-x^2/4\right\}.
\end{align*}
Combining these results,
\begin{align*}
\Pr\nolimits_{H_0}\big\{\big|S&(\sl,t,\el)\big| \geq 2x\big\}\\
&\leq \Pr\nolimits_{H_0}\bigg\{\left|S(\sl,t,\el)-\frac{(t-\sl+1)(\el-t)}{(\el-\sl)^{3/2}}\widehat{U}(\sl,t,\el)\right| \geq x\bigg\}\\
&~~~~~~~~~~~~~~~~~~~~ + \Pr\nolimits_{H_0}\bigg\{\left|\frac{(t-\sl+1)(\el-t)}{(\el-\sl)^{3/2}}\widehat{U}(\sl,t,\el)\right| \geq x\bigg\}\\
&\leq 4\exp\left\{-x^2/4\right\}+2\exp\left\{- x^2 /72\right\}\leq 6\exp\left\{- x^2 /72\right\}.
\end{align*}
By a union bound,
\begin{align*}
\Pr\nolimits_{H_0}\Big\{\max_{\ell \in [L]}\A_{n,\ell}>2x\Big\}\leq n^3\max_{(\sl,t,\el)}\Pr\left\{\left|S(\sl,t,\el)\right| \geq 2x\right\}\leq 6n^3\exp\left\{- x^2 /72\right\}.
\end{align*}
It follows that
\begin{align*}
\Pr\nolimits_{H_0}\Big\{\max_{b\in[B]}\|\g(\pi_b)\|_\infty>2x\Big\}\leq 6Bn^3\exp\left\{- x^2 /72\right\},
\end{align*}
Taking $x=24\sqrt{\log n}$ yields
\[
\Pr\nolimits_{H_0}\left\{\max_{b\in[B]}\|\g(\pi_b)\|_\infty>48\sqrt{\log n}\right\}\leq 6B/n^5,
\]
thus $\thresh_{\alpha,B}< 48\sqrt{\log n}$ with probability at least $1-6B/n^5$.

\section{Additional simulation results}\label{suppsec:simulation}

Table \ref{tab_S:smry} summaries the \art\ methodology and its corresponding applications in numerical experiments presented in the supplementary material.

\begin{table}[!ht]
\small
\centering
\caption{Summary of the \art\ methodology.\label{tab_S:smry}}
\begingroup
\setlength{\tabcolsep}{3pt} 
\renewcommand{\arraystretch}{0.9} 
\begin{threeparttable} 
\begin{tabular}{cccc}
\toprule
Models & Transformation & Aggregation & Result \\
\midrule
\multirow{2}{*}{$\substack{\text{Mean}}$} 
& $K$-means & Rank CUSUM & Figure \ref{suppfig:B}, Figure \ref{suppfig:trans} \\
& $\dev(z;\widehat{f}_\mathcal{D})=-\phi(z-\widehat{\theta}_\mathcal{D})$ & Nonparametric likelihood &  Figure \ref{suppfig:trans} \\
\hline
\multirow{2}{*}{$\substack{\text{Regression}}$} 
& $K$-means & Rank CUSUM & Table \ref{suppfig:trans_reg(i)}, Table \ref{suppfig:trans_reg(ii)}, Table \ref{table_AR} \\

& $\dev(z;\widehat{f}_\mathcal{D})=(y-x^\top\widehat{\theta}_\mathcal{D})^2$ & Nonparametric likelihood & Table \ref{suppfig:trans_reg(i)},  Table \ref{suppfig:trans_reg(ii)} \\
\hline
\multirow{2}{*}{$\substack{\text{Distribution}}$} & $K$-means & Rank CUSUM &  Figure \ref{suppfig:trans}   \\

& $\dev(z;\widehat{f}_\mathcal{D})=-\phi(z)$ & Nonparametric likelihood & Figure \ref{suppfig:trans}   \\
\bottomrule
\end{tabular}
\end{threeparttable}
\endgroup
\end{table}

\subsection{The impact of parameter $B$}\label{supp:B}

In this section, we examine the performance of the \art\ test under different choices of $B$. We focus on AMOC scenarios in the mean change model to illustrate how $B$ affects size and power; analogous findings hold for other models. The model setup and implementation of \art\ follows exactly the procedures described in Section \ref{subsubsec:test}.

\begin{figure}[!ht]
\centering
\includegraphics[width=1\textwidth,height=0.37\textwidth]{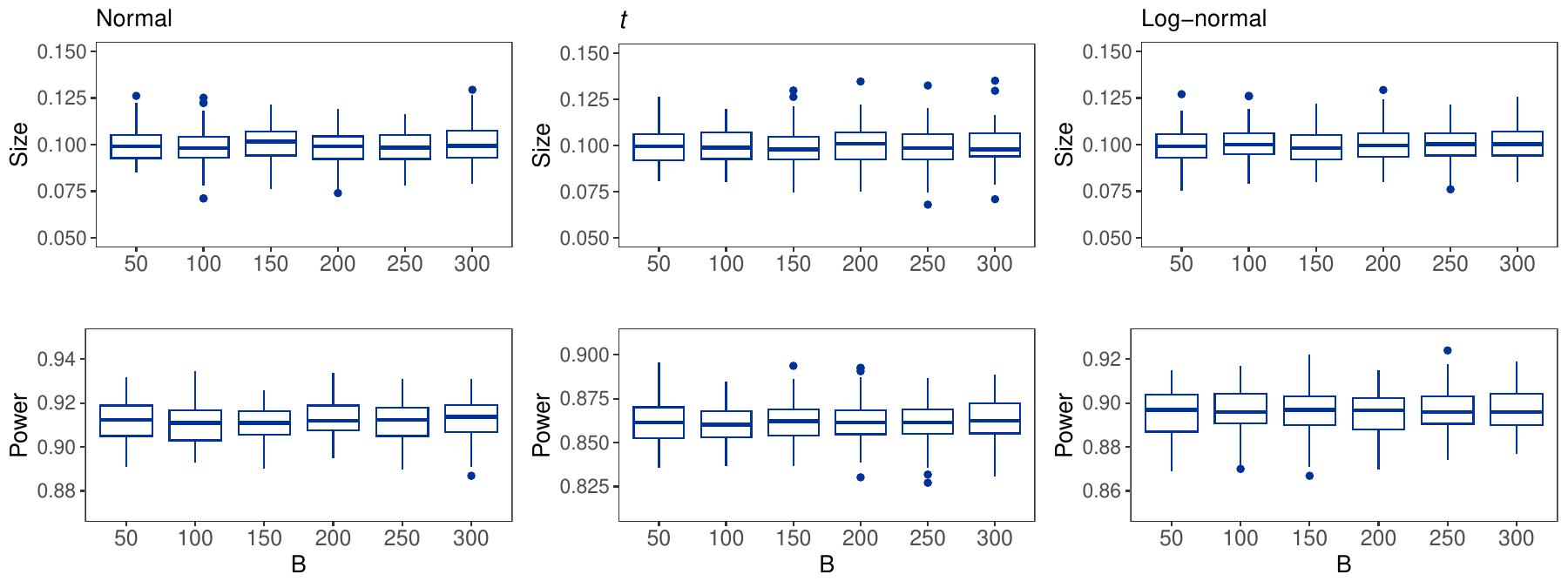}
\caption{Empirical size and power comparisons for varying $B$ under various error distributions, with $\alpha=0.1$ and $(n,d,c_{\theta})=(200,100,0.4)$.}
\label{suppfig:B}
\end{figure}

Figure \ref{suppfig:B} presents boxplots of empirical size and power under various error distributions, based on 100 runs each with 1,000 replications. The results show that the test size is very stable across different values of $B$, consistent with our theoretical findings. The power remains largely robust, with slight improvements observed for larger values of $B$.

\subsection{Deviance versus clustering transformations}

\begin{figure}[!ht]
\centering
\includegraphics[scale=0.6]{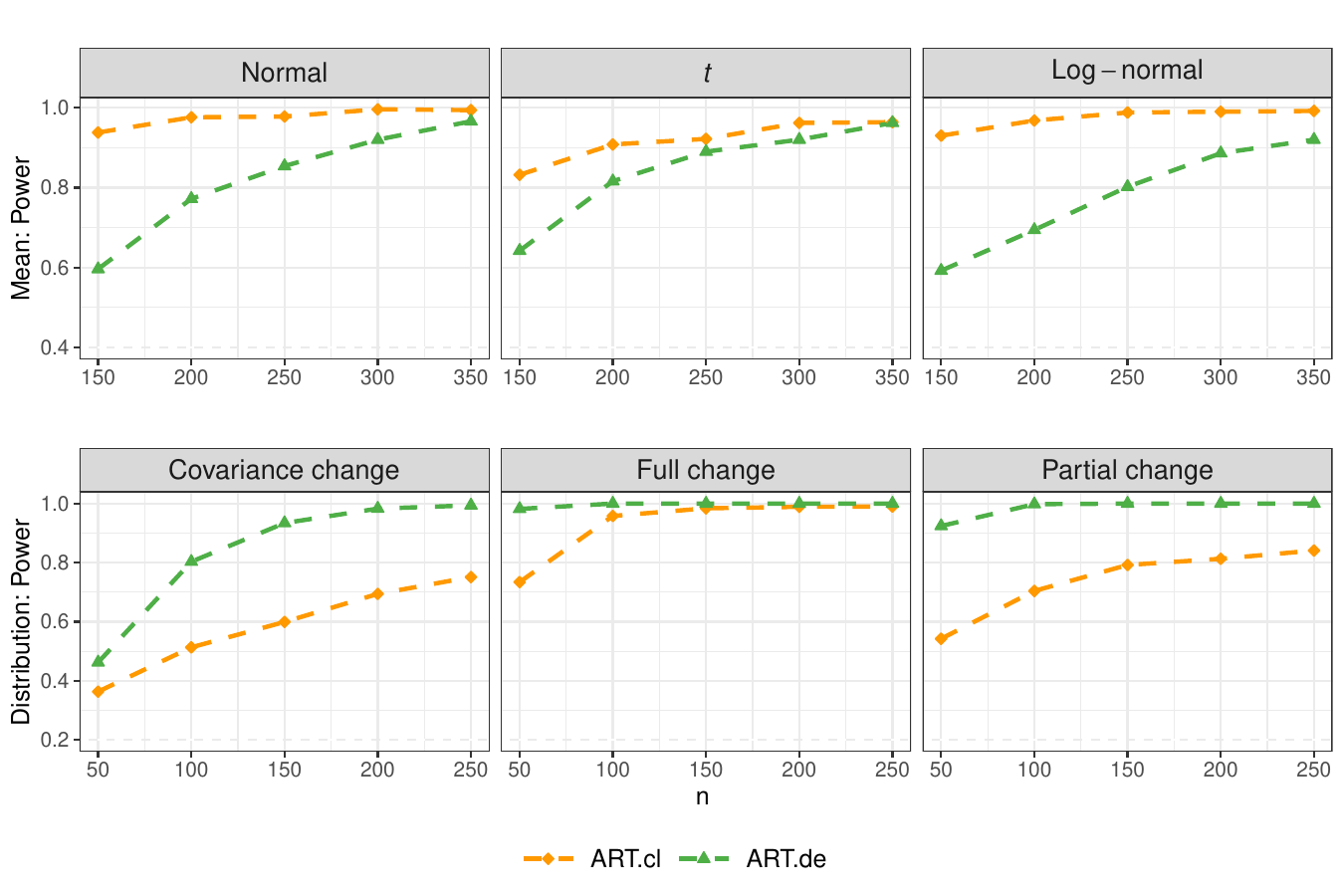}
\caption{Empirical power of the clustering-based (\art.cl) and deviance-based (\art.de) \art\ tests under mean and distributional change models.}
\label{suppfig:trans}
\end{figure}

In the \art\ method, any symmetric transformation ensures exact control of the test size. However, the test power depends on the choice of transformation. Below, we compare the deviance and clustering transformations proposed in Section \ref{subsec:trans}.

We first examine mean changes and distributional changes. The top row of Figure \ref{suppfig:trans} compares power across deviance and clustering transformations while varying $n$ under mean change models. Data are generated as in Section \ref{subsubsec:test} with $d=5$, $s=3$, $\cT^*=\{\lfloor0.3n\rfloor\}$, and $c_P=(0.25,\sqrt{3},1)$ and $c_{\theta}=(0.5,0.8,0.1)$ for normal, $t$, and log-normal errors, respectively.

The second row of Figure \ref{suppfig:trans} illustrates power under distributional changes, investigating three scenarios: (i) Covariance change: $P^*_{1}=\mathcal{N}(0, I/2)$ and $P^*_{2}=\mathcal{N}(\mu_1,(0.9^{|i-j|}))$ with $\mu_1=(0.6,-0.6,0.6,0^{\T}_{d-3})^{\T}$; (ii) Full change: $P^*_{1}=\mathcal{N}(\mu_1,I)$ and $P^*_{2}=\{t(3)\}^d$; and (iii) Partial change: $P^*_{1}=\mathcal{N}(\mu_1,I)$ and $P^*_{2}=\{t(3)\}^s\cdot\{\mathcal{N}(0,1)\}^{d-s}$, where $s=\lfloor0.6d\rfloor$. We set $d=5$ and $\cT^*=\{\lfloor0.3n\rfloor\}$.

\begin{table}[htbp]
\small
\centering
\caption{Empirical size and power of the clustering-based (\art.cl) and deviance-based (\art.de) \art\ tests under various error distributions for the regression model scenario (i).\label{suppfig:trans_reg(i)}}
\begingroup
\setlength{\tabcolsep}{4pt} 
\renewcommand{\arraystretch}{0.9} 
\begin{tabular}{ccccccccccccc}
\toprule
& & &     \multicolumn{2}{c}{Null} && \multicolumn{2}{c}{Small change}  && \multicolumn{2}{c}{Large change} \\ 
\cline {4-5}  \cline {7-8}  \cline{10-11} 
$\tau_1^*$ & Error &$(n, d)$  &    
\art.cl & \art.de  && 
\art.cl & \art.de  && 
\art.cl & \art.de   \\ 
\hline
\multirow{9}{*}{$\lfloor 0.3n \rfloor$} & \multirow{3}{*}{$\substack{\text{Normal}}$}
&(100,200) & 0.097 & 0.102 && 0.640 & 0.566 && 0.791 & 0.875 \\
& &(200,200) & 0.092 & 0.113 && 0.799 & 0.829 && 0.906 & 0.997   \\
& &(400,200) & 0.095 & 0.098 && 0.863 & 0.987 && 0.922 & 1.000 \\
\cline{3-11}
& \multirow{3}{*}{$\substack{t}$}
&(100,200) &0.102 & 0.096 &&  0.387 & 0.368 && 0.549 & 0.656    \\
& &(200,200) &0.089 & 0.101 &&  0.624 & 0.635 && 0.805 & 0.944\\
& &(400,200) & 0.092 & 0.088 &&  0.715 & 0.901 && 0.866 & 1.000 \\
\cline{3-11}
& \multirow{3}{*}{$\substack{\text{Log-normal}}$}
&(100,200) & 0.106 & 0.095 && 0.378 & 0.337 && 0.559 & 0.629   \\
& &(200,200) & 0.111 & 0.091  && 0.574 & 0.585 && 0.787 & 0.914 \\
& &(400,200) & 0.099 & 0.098 && 0.693 & 0.862 && 0.864 & 1.000 \\
\hline
\multirow{9}{*}{$\lfloor 0.4n \rfloor$} & \multirow{3}{*}{$\substack{\text{Normal}}$}
&(100,200) &  0.098 & 0.093  && 0.656 & 0.716 && 0.765 & 0.962 \\
& &(200,200) & 0.095  & 0.098  && 0.791 & 0.916 && 0.877 & 0.999 \\
& &(400,200) & 0.096  & 0.105  && 0.893 & 0.999 && 0.899 & 1.000 \\
\cline{3-11}
& \multirow{3}{*}{$\substack{t}$}
&(100,200) & 0.100 & 0.101 && 0.423 & 0.509 && 0.580 & 0.818   \\
& &(200,200) & 0.087 & 0.102 && 0.625 & 0.773 && 0.795 & 0.985 \\
& &(400,200) & 0.115 & 0.099 && 0.759 & 0.971 && 0.874 & 1.000 \\
\cline{3-11}
& \multirow{3}{*}{$\substack{\text{Log-normal}}$}
&(100,200) & 0.095 & 0.094 && 0.418 & 0.466 && 0.588 & 0.820 \\
& &(200,200) & 0.097 & 0.094 && 0.607 & 0.692 && 0.778 & 0.983 \\
& &(400,200) & 0.093 & 0.104 && 0.752 & 0.933 && 0.876 & 1.000 \\
\hline
\multirow{9}{*}{$\lfloor 0.5n \rfloor$} & \multirow{3}{*}{$\substack{\text{Normal}}$}
&(100,200) &  0.093 & 0.099  && 0.642 & 0.751 && 0.748 & 0.960 \\
& &(200,200) & 0.106 & 0.105  && 0.823 & 0.956 && 0.878 & 1.000 \\
& &(400,200) & 0.100 & 0.115  && 0.922 & 1.000 && 0.911 & 1.000 \\
\cline{3-11}
&  \multirow{3}{*}{$\substack{t}$}
&(100,200) & 0.098 & 0.101 &&  0.399 & 0.579 && 0.575 & 0.878 \\
& &(200,200) & 0.097 &0.108  &&  0.622 & 0.829 && 0.758 & 0.993 \\
& &(400,200) & 0.103 & 0.099 &&  0.789 & 0.989 && 0.870 & 1.000  \\
\cline{3-11}
& \multirow{3}{*}{$\substack{\text{Log-normal}}$}
&(100,200) & 0.089 & 0.103 && 0.388 & 0.506 && 0.608 & 0.858   \\
& &(200,200) & 0.106 & 0.101 && 0.610 & 0.805 && 0.772 & 0.988 \\
& &(400,200) & 0.097 & 0.098 && 0.812 & 0.961 && 0.874 & 1.000  \\     
\bottomrule
\end{tabular}
\endgroup
\end{table}

\begin{table}[htbp]
\small
\centering
\caption{Empirical size and power of the clustering-based (\art.cl) and deviance-based (\art.de) \art\ tests under various error distributions for the regression model scenario (ii).\label{suppfig:trans_reg(ii)}}
\begingroup
\setlength{\tabcolsep}{4pt} 
\renewcommand{\arraystretch}{0.9} 
\begin{tabular}{ccccccccccccc}
\toprule
& & &     \multicolumn{2}{c}{Null} && \multicolumn{2}{c}{Small change}  && \multicolumn{2}{c}{Large change} \\ 
\cline {4-5}  \cline {7-8}  \cline{10-11} 
$\tau_1^*$ & Error &$(n, d)$  &    
\art.cl & \art.de  && 
\art.cl & \art.de  && 
\art.cl & \art.de   \\ 
\hline
\multirow{9}{*}{$\lfloor 0.3n\rfloor$} & \multirow{3}{*}{$\substack{\text{Normal}}$}
&(100,200) & 0.103 & 0.103  && 0.714 & 0.226 && 0.729 & 0.541 \\
& &(200,200) & 0.094 & 0.097  && 0.856 & 0.379 && 0.878 & 0.825 \\
& &(400,200) & 0.114 & 0.092 && 0.848 & 0.595 && 0.975 & 0.991\\
\cline{3-11}
& \multirow{3}{*}{$\substack{t}$}
&(100,200)  & 0.099 & 0.101 && 0.450 & 0.186 && 0.520 & 0.335      \\
& &(200,200) & 0.108 & 0.092 && 0.738 & 0.264 && 0.798 & 0.644 \\
& &(400,200) & 0.098 & 0.109 && 0.914 & 0.427 && 0.933 & 0.921 \\
\cline{3-11}
& \multirow{3}{*}{$\substack{\text{Log-normal}}$}
&(100,200) & 0.107 & 0.096 && 0.410 & 0.162 && 0.497 & 0.361      \\
& &(200,200) & 0.108 & 0.099 && 0.684 & 0.256 && 0.789 & 0.619      \\
& &(400,200) & 0.089 & 0.092 && 0.852 & 0.417 && 0.884 & 0.887     \\
\hline
\multirow{9}{*}{$\lfloor 0.4n\rfloor$} & \multirow{3}{*}{$\substack{\text{Normal}}$}
&(100,200) &  0.101 & 0.097  && 0.763 & 0.297 && 0.766 & 0.730    \\
& &(200,200) & 0.089  & 0.098  && 0.876 & 0.484 && 0.922 & 0.952   \\
& &(400,200) & 0.097  & 0.111  && 0.923 & 0.782 && 0.977 & 1.000  \\
\cline{3-11}
& \multirow{3}{*}{$\substack{t}$}
&(100,200) & 0.099 & 0.101 && 0.518 & 0.237 && 0.571 & 0.579       \\
& &(200,200) & 0.108 & 0.110 && 0.812 & 0.382 && 0.829 & 0.852    \\
& &(400,200) & 0.107 & 0.101 && 0.946 & 0.575 && 0.959 & 0.988    \\
\cline{3-11}
& \multirow{3}{*}{$\substack{\text{Log-normal}}$}
&(100,200) & 0.097 & 0.100 && 0.511 & 0.226 && 0.596 & 0.501     \\
& &(200,200) & 0.086 & 0.092 && 0.833 &0.363  && 0.850 & 0.814   \\
& &(400,200) & 0.089 & 0.099 && 0.922 & 0.569 && 0.931 & 0.980   \\
\hline
\multirow{9}{*}{$\lfloor 0.5n\rfloor$} & \multirow{3}{*}{$\substack{\text{Normal}}$}
&(100,200) & 0.110  & 0.108  && 0.730 & 0.337 && 0.803 & 0.772       \\
& &(200,200) & 0.096  & 0.107  && 0.916 & 0.555 && 0.941 & 0.964       \\
& &(400,200) &  0.101 & 0.098  && 0.979 & 0.832 && 0.989 & 0.999      \\
\cline{3-11}
&  \multirow{3}{*}{$\substack{t}$}
&(100,200) &  0.103&  0.099 && 0.505 & 0.279 && 0.623 & 0.660  \\
& &(200,200) & 0.091  &0.108  && 0.844 & 0.474 && 0.906 & 0.890    \\
& &(400,200) & 0.097 & 0.099 && 0.965 & 0.676 && 0.966 & 0.995     \\
\cline{3-11}
& \multirow{3}{*}{$\substack{\text{Log-normal}}$}
&(100,200) & 0.104 & 0.097 && 0.508 & 0.236 && 0.601 & 0.636   \\
& &(200,200) & 0.094 & 0.103 && 0.804 & 0.396 && 0.902 & 0.868    \\
& &(400,200) & 0.099 & 0.094  && 0.968 & 0.629 && 0.982 & 0.991  \\     
\bottomrule
\end{tabular}
\endgroup
\end{table}

Next, we consider changes in regression coefficients under two scenarios:
\begin{itemize}
    \item[(i)] $\theta_1^*=(0.5,0,-0.5,0_{d-3}^{\T})^{\T}$ and $\theta_{2}^*=\theta_1^* + c_{\theta} \cdot D_{1,s}$ with $D_{1,s}=(-1,1,-1,1,-1,0_{d-5}^{\T})^{\T}$;
    \item[(ii)] $\theta_1^*=(0.5,-0.5,-0.5,0.5,0_{d-4}^{\T})^{\T}$ and $\theta_2^*=\theta_1^* + c_{\theta} \cdot D_{1,s}$. 
\end{itemize}
We use the same three error distributions as in Section \ref{subsubsec:test}. Tables \ref{suppfig:trans_reg(i)}--\ref{suppfig:trans_reg(ii)} display empirical size and power for $\tau^*_1=\{\lfloor0.3n\rfloor,\lfloor0.4n\rfloor,\lfloor0.5n\rfloor\}$ and $c_{\theta}=\{0.5,0.7\}$ corresponding to small and large changes.

Both clustering and deviance transformations exhibit satisfactory performance, with power increasing as sample size or the magnitude of change grows. Clustering transformations can incorporate diverse clustering algorithms, making them adaptable to complex data. Meanwhile, deviance transformations are computationally efficient and straightforward to implement. Both approaches each have their own merits, rendering them reliable options for a wide range of practical applications.

\subsection{Departures from independence}

We assess \art\ when the data $\{Z_i\}_{i=1}^n$ are not independent. Specifically, we consider regression models with an autoregressive (AR) process among $\{\varepsilon_i\}_{i=1}^n$: $\varepsilon_i=0.6 \varepsilon_{i-1}+\sqrt{1-0.6^2} e_i$, where $e_i$ follows one of the three error distributions $P_{\epsilon,1}$ described in Section \ref{subsec:simu}. The model settings aligns with those in Section \ref{subsec:simu}, focusing on changes in regression coefficients for both small and large shifts ($c_{\theta}=0.5$ and $c_{\theta}=1$).

Table \ref{table_AR} compares the performance of \art\ under independent and such AR(1) dependent scenarios, with $\alpha=0.1$. The empirical size remains near the nominal level in both cases, while the power under the AR(1) setting is comparable to that under independence. These outcomes highlight the robustness and effectiveness of \art\ for certain nonexchangeable data structures, meriting further theoretical investigation.

\begin{table}[htbp]
\small
\centering
\caption{Empirical size and power of \art\ under independent and AR(1) error structures for various error distributions in regression models.\label{table_AR}}
\begingroup
\setlength{\tabcolsep}{3pt} 
\renewcommand{\arraystretch}{0.9} 
\begin{tabular}{cccccccccccccc}
\toprule
& &     \multicolumn{2}{c}{Null} && \multicolumn{2}{c}{Small change}  && \multicolumn{2}{c}{Large change} \\ 
\cline {3-4}  \cline {6-7}  \cline{9-10}
Error &$(n, d)$  &    
Independent & AR(1) && 
Independent & AR(1) && 
Independent & AR(1) \\ \hline
\multirow{6}{*}{$\substack{\text{Normal}}$}
&(50,100)  & 0.096 & 0.110 && 0.401 & 0.402 && 0.445 & 0.443 \\
&(100,100) & 0.101 & 0.104 && 0.581 & 0.594 && 0.765 & 0.761 \\
&(200,100) & 0.091 & 0.111 && 0.621 & 0.654 && 0.946 & 0.941 \\
&(50,200)  & 0.102 & 0.107 && 0.259 & 0.284 && 0.285 & 0.334 \\
&(100,200) & 0.108 & 0.112 && 0.458 & 0.490 && 0.753 & 0.733 \\
&(200,200) & 0.097 & 0.105 && 0.601 & 0.571 && 0.916 & 0.932 \\
\hline
\multirow{6}{*}{$\substack{t}$}
&(50,100)  & 0.094 & 0.110 && 0.301 & 0.299 && 0.362 & 0.335 \\
&(100,100) & 0.106 & 0.114 && 0.405 & 0.403 && 0.645 & 0.636  \\
&(200,100) & 0.099 & 0.106 && 0.463 & 0.438 && 0.902 & 0.901  \\
&(50,200)  & 0.097 & 0.106 && 0.205 & 0.194 && 0.259 & 0.245 \\
&(100,200) & 0.092 & 0.120 && 0.293 & 0.356 && 0.613 & 0.616 \\
&(200,200) & 0.093 & 0.123 && 0.368 & 0.394 && 0.899 & 0.879 \\
\hline
\multirow{6}{*}{$\substack{\text{Log-normal}}$}
&(50,100)  & 0.104 & 0.112 && 0.231 & 0.209 && 0.309 & 0.314 \\
&(100,100) & 0.101 & 0.105 && 0.382 & 0.345 && 0.619 & 0.618 \\
&(200,100) & 0.105 & 0.098 && 0.395 & 0.404 && 0.896 & 0.896  \\
&(50,200)  & 0.098 & 0.104 && 0.187 & 0.159 && 0.218 & 0.236\\
&(100,200) & 0.094 & 0.113 && 0.278 & 0.282 && 0.593 & 0.611 \\
&(200,200) & 0.105 & 0.112 && 0.329 & 0.352 && 0.876 & 0.880 \\
\bottomrule
\end{tabular}
\endgroup
\end{table}

\end{document}